\documentclass[sigconf]{acmart}
\AtBeginDocument{%
  \providecommand\BibTeX{{%
    \normalfont B\kern-0.5em{\scshape i\kern-0.25em b}\kern-0.8em\TeX}}}

\copyrightyear{2024}
\acmYear{2024}
\setcopyright{rightsretained}
\acmConference[UIST '24]{The 37th Annual ACM Symposium on User Interface Software and Technology}{October 13--16, 2024}{Pittsburgh, PA, USA}
\acmBooktitle{The 37th Annual ACM Symposium on User Interface Software and Technology (UIST '24), October 13--16, 2024, Pittsburgh, PA, USA}
\acmDOI{10.1145/3654777.3676414}
\acmISBN{979-8-4007-0628-8/24/10}

\usepackage{enumitem}
\usepackage{hyperref}
\usepackage{tabularx}

\newcommand{\toolname}{VizAbility}

\newcommand{\boxedtext}[1]{\mbox{\scriptsize\bfseries\textsf{#1}}} 
\newcommand{\nota}[2]{
	\boxedtext{#1}
	{\small$\blacktriangleright$\textsl{#2}}
}

\newcommand\nam[1]{{ \color{magenta}\nota{Nam}{#1}}}

\newcommand\change[1]{{ \color{blue} {#1} }}

\usepackage{soul}
\usepackage{xcolor}
\usepackage[scale=0.82]{FiraMono}

\definecolor{maroon}{RGB}{138, 16, 11}
\definecolor{gray}{RGB}{114, 97, 88}
\definecolor{gold}{RGB}{220, 202, 160}

\newcommand{\hlmaroon}[1]{\colorbox{maroon}{\textcolor{white}{#1}}}

\newcommand{\hlgold}[1]{\colorbox{gold}{\textcolor{black}{#1}}}
            
\begin{document}

\title{\toolname{}: Enhancing Chart Accessibility with LLM-based Conversational Interaction}

\author{Joshua Gorniak}
\email{joshua.gorniak@bc.edu}
\affiliation{%
  \institution{Boston College}
  \streetaddress{140 Commonwealth Ave}
  \city{Chestnut Hill}
  \state{Massachusetts}
  \country{USA}
  \postcode{02467}
}

\author{Yoon Kim}
\email{yoonkim@mit.edu}
\affiliation{%
  \institution{MIT}
  \streetaddress{77 Massachusetts Ave}
  \city{Cambridge}
  \state{Massachusetts}
  \country{USA}
  \postcode{02139}
}

\author{Donglai Wei}
\email{donglai.wei@bc.edu}
\affiliation{%
  \institution{Boston College}
  \streetaddress{140 Commonwealth Ave}
  \city{Chestnut Hill}
  \state{Massachusetts}
  \country{USA}
  \postcode{02467}
}

\author{Nam Wook Kim}
\email{nam.wook.kim@bcu}
\affiliation{%
  \institution{Boston College}
  \streetaddress{140 Commonwealth Ave}
  \city{Chestnut Hill}
  \state{Massachusetts}
  \country{USA}
  \postcode{02467}
}

\renewcommand{\shortauthors}{Gorniak et al.}

\begin{abstract}
Traditional accessibility methods like alternative text and data tables typically underrepresent data visualization's full potential. Keyboard-based chart navigation has emerged as a potential solution, yet efficient data exploration remains challenging. We present \toolname{}, a novel system that enriches chart content navigation with conversational interaction, enabling users to use natural language for querying visual data trends. \toolname{} adapts to the user's navigation context for improved response accuracy and facilitates verbal command-based chart navigation. Furthermore, it can address queries for contextual information, designed to address the needs of visually impaired users. We designed a large language model (LLM)-based pipeline to address these user queries, leveraging chart data \& encoding, user context, and external web knowledge. We conducted both qualitative and quantitative studies to evaluate \toolname{}'s multimodal approach. We discuss further opportunities based on the results, including improved benchmark testing, incorporation of vision models, and integration with visualization workflows.
 

\end{abstract}

\ccsdesc[500]{Human-centered computing~Interactive systems and tools}
\ccsdesc[500]{Human-centered computing~Visualization systems and tools}

\keywords{data visualization, accessibility, blind and low vision people}

\begin{teaserfigure}
    \centering
  \includegraphics[width=\textwidth]{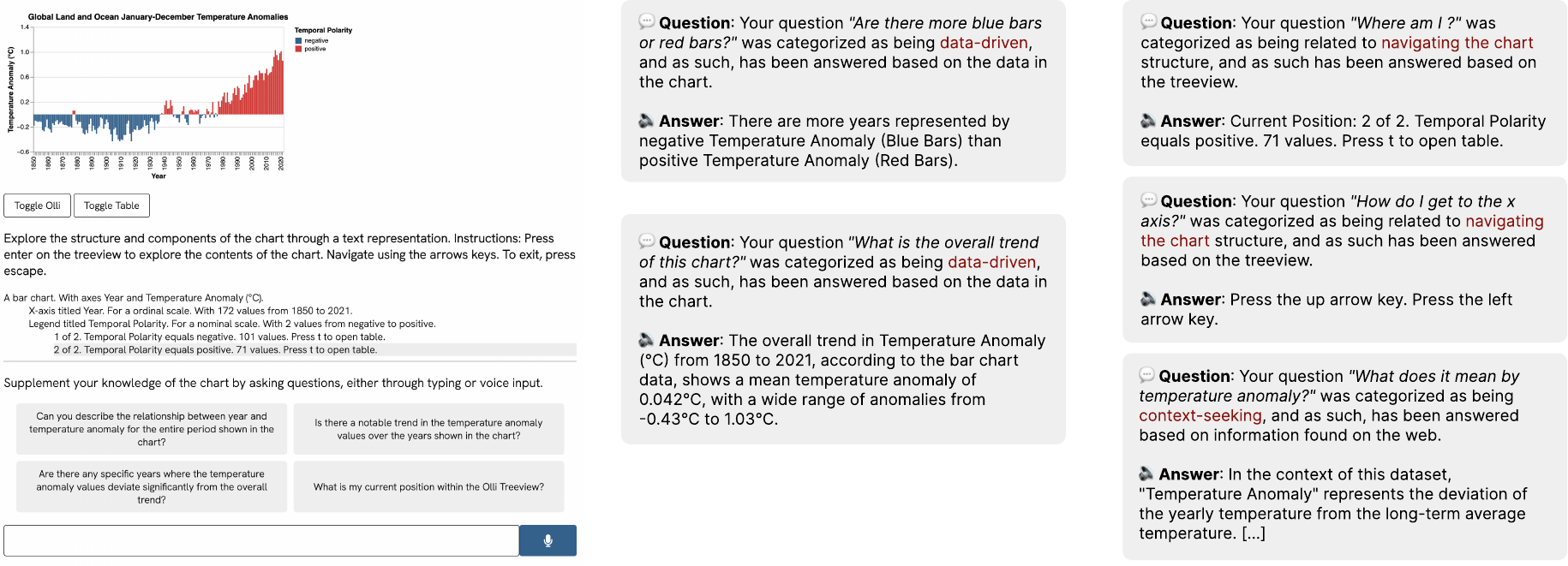}
  \caption{\toolname{}'s overall user interface presents the keyboard navigation of chart content or data tables augmented with the ability to ask natural language questions. Example queries are shown on the right side, including visual \& data queries, navigation, and context-seeking questions.}
  \Description{An overview of \toolname{}'s user interface. A bar chart is displayed along with a text representation of its contents. The interface for the conversational module is depicted below it. Off to the right lies a condensed overview of the pipeline for answering user questions via the conversational model.}
  \label{fig:teaser}
\end{teaserfigure}


\maketitle

\section{Introduction}

Data visualization has become an indispensable tool in our broader society, aiding in the comprehension of important information and facilitating informed decision-making~\cite{lee2020reaching}. Its strength stems from leveraging the vast information bandwidth of our visual perception, which surpasses other sensory modalities~\cite{fekete2008value}.
However, an over-reliance on visual representation can inadvertently marginalize those with blindness or low vision (BLV), restricting their ability to engage with and understand data visualizations~\cite{marriott2021inclusive}. Individuals with BLV often come across data visualizations while using screen readers such as JAWS, NVDA, and VoiceOver to navigate the web~\cite{kim2021accessible,siu2021covid}. Unfortunately, a significant portion of data visualizations on the web remains largely inaccessible to this group~\cite{siu2021covid,joyner2022visualization}, resulting in a pronounced information gap.

Numerous assistive technologies have been developed to allow BLV users to access visualizations using sensory modalities other than vision~\cite{kim2021accessible}.
Tactile visualizations can provide a tangible representation of data while necessitating specialized hardware such as haptic displays ~\cite{paneels2009review} and embossing machines~\cite{engel2017analysis}. On the other hand, sonification can enable users to discern trends and anomalies through sound~\cite{wang2022seeing}, but it is typically limited to single-series data.
Traditional methods for adapting web visualizations for screen readers, such as data tables and alternative text~\cite{kim2021accessible}, often compromise the unique benefits of visualizations. Keyboard-based chart navigation~\cite{weninger2015asvg,elavsky2023data,thompson2023chart,zong2022rich} have emerged as an alternative solution. However, orienting oneself and navigating within complex chart encoding structures pose challenges for efficient data exploration~\cite{kim2023beyond}.

This work introduces \toolname{}, a novel approach to augment keyboard navigation of chart content with conversational interaction (\autoref{fig:teaser}). First, we use Olli's tree view (referenced in \cite{2022-olli}) to create a keyboard-navigable text representation of a chart. Next, we enhance this tree view with an LLM-based question-and-answer module for addressing on-demand queries. These natural language queries allow users to understand the chart without needing to navigate and mentally synthesize different parts of the chart to derive insights. \toolname{} uses the user's position within the tree view to efficiently respond to both \textsc{visual} and \textsc{analytical} queries, facilitating the exploration of data trends and visual patterns. Moreover, \toolname{} can handle \textsc{contextual} queries, providing background information about the chart, tailored specifically for the needs of BLV individuals~\cite{kim2023exploring}. We based these query types on actual questions asked by BLV people in previous studies~\cite{kim2020answering, kim2023exploring}. Additionally, it can manage \textsc{navigation} queries, allowing users to control their position in the tree view through verbal commands. 






Our LLM-based pipeline first uses few-shot prompting to classify user queries into \textsc{visual}, \textsc{analytical}, \textsc{contextual}, and \textsc{navigation} queries. Once classified, \toolname{} employs a query-specific prompting strategy. For analytical and visual queries, we aggregate both the chart's transformed data and color encoding into one CSV file, which is subsequently fed along with the keyboard-navigable text representation with the user's location~\cite{2022-olli} to the LLM via a CSV Agent~\cite{csvagent}. \textsc{Contextual} queries utilize a Web Browser Agent~\cite{serpapi}, whereas \textsc{navigation} queries employ the LLM to discern the starting/ending nodes from a user query and employ a breadth-search algorithm to calculate the shortest path between the nodes. We designed the prompts to minimize hallucinations and address unanswerable queries via structured output formatting. We collaborated with a blind participant in the development of \toolname{}, holding a series of feedback sessions.





We carried out quantitative assessments to evaluate the question \& answering pipeline. We evaluated response quality using a combined dataset of 979 real BLV user questions derived from previous research~\cite{kim2023exploring} and 48 synthetically generated navigation queries. Splitting the dataset, 80\% was used for testing and 20\% for validation. Our query classification achieved an accuracy of 87.39\%. The final response evaluation involved a manual assessment by two researchers, followed by a more scalable LLM-based evaluation using GPT4.
Both the human and GPT4 assessments followed a 5-point Likert Scale that assigned a value ranging from ``Very Poor'' to ``Very Good'' depending on the response's coherence to the ground truth. For the human evaluation, $77.36\%$ and $5.14\%$ of the responses were rated as ``Very Good' and ``Good'' respectively, contributing to an overall rate of $82.50\%$. We computed Kendall's $\tau$ score to assess the consistency between the human and GPT4 assessment methods, observing significant alignment ($=0.5526$,  $p<0.001$). As a baseline comparison, we used the GPT4-based evaluation to assess responses generated by GPT-4 with vision (GPT-4V) on the same test data. We observed the performance of GPT-4V was poorer, with only 27.96\% and 10.10\% of responses being designated as ``Very Good'' and ``Good'' respectively.


We conducted a preliminary usability study with six BLV participants recruited through the National Institute for the Blind. Initially, participants explored \toolname{} without guidance and were subsequently introduced to various query types. They also completed the System Usability Scale survey. The results suggest that while participants could learn to use the system, discerning query types without guidance proved challenging. Nonetheless, they acknowledged the merits of the integrated approach and offered suggestions for further improvements and potential applications. For instance, we introduced data tables as an alternative to the tree view and added cold-start query recommendations to assist users in getting started. Combining insights from both quantitative and qualitative evaluations, we identify potential avenues for future work. These include enhancing user-driven customization, developing a more robust benchmarking system, incorporating vision models, and integrating our solution into existing visualization tools.

Our main contributions lie in the following:
\begin{itemize}
    \item Design and development of \toolname{}, which incorporates an LLM-based question-and-answer module to enhance keyboard navigation of chart content for screen reader users.
    \item Development of a benchmark dataset comprising ground truths and chart specifications that augment existing questions posed by blind individuals~\cite{kim2023exploring}, facilitating further research in this area.
    \item Preliminary evaluation that demonstrates the effectiveness of \toolname{} in comparison to baseline question-and-answer systems, as well as its usability as assessed by blind participants.
\end{itemize}

The \toolname{} source code and dataset are available at \href{https://dwr.bc.edu/vizability/}{https://\-dwr.bc.edu/vizability/}.






\section{Related Work}
\subsection{Accessibility Systems for Data Visualization}


The recent survey offers an overview of previous efforts exploring the use of non-visual modalities, such as speech, sound, and touch~\cite{kim2021accessible}. For example, \textbf{sonification} employs non-speech auditory channels, such as pitch and volume, to represent data~\cite{wang2022seeing,sakhardande2019comparing}. While this can offer users a swift overview of a graph, it struggles to communicate exact values and might not be effective beyond single-series charts~\cite{ferres2013evaluating}. An empirical study indicates that blind individuals favor speech over sonification, as the cognitive load for a sonified graph feels subjectively more intense~\cite{sakhardande2019comparing}.

\textbf{Tactile systems} employ methods like embossed prints, haptic feedback through vibrations, and braille for text representation. These systems enable both simultaneous and on-demand exploration of data trends, offering an advantage over linear audio~\cite{engel2018user}. However, they also necessitate enhanced perceptual motor skills. Similar to sonification, accurately discerning complex structures can be challenging, often demanding a more refined spatial resolution~\cite{engel2017analysis}. Producing tactile graphs typically involves specialized hardware, such as embossers, which might not be economically feasible for the average user~\cite{kim2021accessible}; thus, they are typically used and created in the field of education by teachers~\cite{engel2017improve}.

\textbf{Screen readers}, utilizing text/speech modalities, stand as the predominant assistive technology, particularly for navigating web content. The go-to accessibility techniques for screen readers encompass alternative text and data tables. Yet, these strategies often reduce data visualizations to brief descriptions or mere numbers, undermining their inherent advantages. An alternative approach involves crafting keyboard-navigable text descriptions derived from the chart's structure. A select group of data visualization tools and toolkits, such as HighCharts~\cite{highcharts2022} and amCharts~\cite{amcharts}, offer some degree of this navigation and customization~\cite{kim2023beyond}. In recent times, several research efforts have advanced navigation capabilities, representing charts as traversable graph structures~\cite{weninger2015asvg,godfrey2018accessible,zong2022rich,thompson2023chart,elavsky2023data,seechart}.

\textbf{Voice-based virtual assistants} are emerging as valuable accessibility tools in human-computer interaction~\cite{vtyurina2019verse}. However, only a handful of studies have delved into using natural language interactions for accessing data visualization content. For instance, Murillo-Morales \& Miesenberger~\cite{murillo2017non} showcased a prototype system where users can ask predefined questions related to data metrics such as mean, extremes, and range. In a similar vein, VoxLens~\cite{sharif2022voxlens} facilitates voice-activated interactions capable of addressing basic queries with terms like ``maximum'' and ``median''. Additionally, Kim et al.~\cite{kim2023exploring} used a Wizard-of-Oz approach to study the types of questions blind individuals pose about charts.

To address the limitations of relying on a single sensory modality, \textbf{multi-sensory perception} is frequently utilized. A prevalent strategy involves merging verbal (speech) cues with non-verbal ones, such as sonification, tactile graphics, and haptic feedback. Examples include offering on-demand audio descriptions of touched elements~\cite{goncu2011gravvitas,gardner2008making,landau2001development} or pairing sonification with screen readers~\cite{thompson2023chart,taibbi2014audiofunctions} and braille~\cite{maidr}. However, these solutions often necessitate specialized software and hardware, especially for interactive tactile support, making them expensive to implement.

In this study, we adopt an integrated approach that merges structured chart and table navigation using the keyboard with conversational interaction via verbal commands. Our work builds on the prior work~\cite{kim2023beyond} that suggests the respective advantages of data tables---familiarity, structured chart navigation---deeper engagement, and conversational interaction via natural language commands---faster data exploration. Our primary technical advancement centers on employing LLMs to enhance the current chart question-and-answer mechanism for the visually impaired.

\subsection{Question \& Answering Systems for Data Visualization}

Within the realm of image understanding research, \textbf{visual question answering} has been rigorously explored in both natural language processing and computer vision, specifically regarding answering text-based queries about images~\cite{antol2015vqa,wu2017visual,kafle2017visual}. Yet, the majority of these endeavors have centered on natural scene images rather than human-generated visuals such as data visualizations.

Recent studies have begun to focus on \textbf{data visualization images}~\cite{hoque2022chart}. For example, FigureQA~\cite{kahou2017figureqa} offers a corpus tailored for yes/no questions, such as ``Is Light Gold less than Periwinkle?''. Conversely, DVQA~\cite{kafle2018dvqa} expands its purview to encompass questions about chart structure (``are the bars horizontal?''), data retrieval (``what percent of people prefer A?''), and reasoning (``Is A preferred more than B?''). While both FigureQA and DVQA rely on synthetically generated charts, PlotQA introduces a large-scale dataset of real-world scientific plots. Unlike the templated questions of the aforementioned datasets, ChartQA delivers human-composed questions, enhanced using LLMs~\cite{masry2022chartqa}. These models predominantly process pixel images as input. For instance, they extract data tables and other image features~\cite{masry2022chartqa,kantharaj2022opencqa}, feeding them into vision and language task models~\cite{cho2021unifying}. Consequently, their accuracy largely hinges on their image processing capabilities, often leading to suboptimal results (e.g., failing to recover data values due to the absence of data labels). In a different approach, Kim et al.\cite{kim2020answering} proposed a system that not only answers questions but also provides explanations, operating on Vega-lite\cite{satyanarayan2016vega} instead of images. All the current question-answering systems are limited to basic visualization types like bar, line, and pie charts.

While chart QA systems hint at the potential for enhancing visualization accessibility, they often overlook the \textbf{specific needs of BLV users}. Recent studies have shown that BLV users frame questions differently compared to those with sight~\cite{gurari2018vizwiz,dognin2020image}. A limited number of systems directly address the challenge of crafting question-and-answer systems tailored for the blind~\cite{sharif2022voxlens,murillo2017non}. However, these systems do not always offer specialized features for the blind and are constrained in their question-answering capabilities. For instance, VoxLens~\cite{sharif2022voxlens} is limited to charts with single series data, while the system by Murillo-Morales \& Miesenberger~\cite{murillo2017non} is restricted to bar charts. Kim et al.~\cite{kim2023exploring} have recently curated a set of questions posed by blind individuals through a wizard-of-oz study, laying the groundwork for more refined and targeted question-and-answer systems.

In this paper, we present an enhanced chart question-and-answer system tailored for blind users, leveraging the power of LLMs. Our approach focuses on reasoning, predicated on the availability of chart encoding and underlying data. We utilize Vega-lite as input, thereby accommodating a variety of chart types. The system handles a wide array of queries, including data and visual queries found in existing question-answer systems, as well as contextual and navigation queries specific to chart accessibility.


\section{\toolname{} Design Decisions}
\label{sec:design-decisions}
In this section, we outline the key design decisions made \textit{during the development and evaluation} of \toolname{}. These decisions were guided by a combination of prior empirical research findings and practical considerations based on user feedback throughout the design and development process. \hlmaroon{D}: Decisions relevant to our primary contributions \hlgold{D}: User interface and usability decisions.

\paragraph{\protect\hlgold{D1}: Enable understanding the chart encoding structure}
Bridging the perceptual gap between BLV and sighted individuals requires a deep understanding of what the chart looks like. 
Previous research indicates that navigating charts based on their visual encoding helps BLV users gain a clearer visual understanding~\cite{kim2023beyond}. In this work, we use Olli~\cite{2022-olli} to generate keyboard-navigable text representations of charts. 
This approach provides the visual structure of a chart in symbolic text form, helping answer visually oriented questions. Other keyboard-navigable text representations (e.g., a similar text structure extracted from HighCharts) would provide a similar benefit.


\paragraph{\protect\hlmaroon{D2}: Support efficient data exploration via natural language interaction}

 Furthermore, extracting aggregate measures and discerning perceptual patterns beyond basic value retrievals becomes challenging when navigating data points individually via keyboard input~\cite{kim2023beyond}. This issue exacerbates as the hierarchical text representation becomes deeper and wider with complex chart encodings, resulting in hard mental operations~\cite{green1996usability}. In this work, we adopt a conversational interaction approach that transcends traditional methods such as sonification and tactile perception, which are limited in scalability for modern data visualizations. Leveraging LLMs and user context within keyboard navigation, we address visual and analytic queries that facilitate rapid exploration of nuanced trends and patterns in charts.

\paragraph{\protect\hlmaroon{D3}: Provide contextual knowledge on demand for better chart comprehension} 
Current chart question and answering systems often neglect the distinct types of questions posed by blind versus sighted individuals. Recent research involving blind participants indicates that they frequently ask contextual questions alongside data-related and visual inquiries~\cite{kim2023exploring}. These questions often seek external information not present in the chart, such as meanings about axes or specific data labels. Providing answers to these inquiries can enhance the self-efficacy and autonomy of blind individuals. In our approach, we use an LLM with web search capabilities to address these contextual queries.

\paragraph{\protect\hlmaroon{D4}: Alleviate the difficulty of keyboard-based chart navigation}

Navigating complex chart structures can become less intuitive and more cumbersome~\cite{kim2023beyond,zong2022rich}, when restricted to keyboard inputs alone. In our work, we aim to mitigate this challenge by facilitating nonlinear investigation across the chart structure via speech commands. In our work, we address this challenge by enabling nonlinear exploration of chart structures through speech commands. This multimodal approach enhances flexibility and efficiency. Furthermore, we aim to assist users in orienting themselves within the chart structure. Understanding one's position within a digital chart holds equal importance to spatial awareness in physical mobility~\cite{chundury2021towards}.

\paragraph{\protect\hlgold{D5}: Provide a fallback strategy using familiar data presentation format}
The hierarchical text representation of charts, while effective, can sometimes be seen as overly complex for certain users. This observation was noted in previous research~\cite{kim2023beyond}, as well as in our user study (see Section \ref{sec:user-study}). In response, we offer conventional data tables as an alternative to navigating the chart structure. This option is advantageous due to its compatibility with screen readers and widespread user familiarity~\cite{zong2022rich,kim2023beyond}. Additionally, we incorporate the user's context within the data table to enhance our system's ability to accurately respond to data-oriented queries.

\paragraph{\protect\hlmaroon{D6}: Implement error prevention strategies for enhanced LLM interaction}
User queries can often be ambiguous or not directly related to the available data. Likewise, LLMs can face technical limitations, such as time-outs or processing errors, which can disrupt the interaction flow. Strategies to anticipate and mitigate these issues can help manage user expectations and offer a fluid user experience. We address these challenges by implementing \textit{proactive} query refinement for ambiguous queries and \textit{reactive} suggestions of alternative queries in case of failures.

\paragraph{\protect\hlgold{D7}: Follow standard accessibility principles}
The W3C Web Accessibility Initiative specifies four essential principles for web accessibility: \textbf{P}erceivable, \textbf{O}perable, \textbf{U}nderstandable, and \textbf{R}obust~\cite{waiprinciples}, while Chartability adds three more for visualization:
\textbf{C}ompromising, 
\textbf{A}ssist-ive, and 
\textbf{F}lexible~\cite{chartability_pour_caf}. Ensuring these principles in new assistive technology can be often overlooked. We strive to adhere to them in designing \toolname{}. Examples include reaffirming user queries (P); suggesting cold-start queries (O); indicating delays in responses (U); allowing both speech and text inputs to formulate queries (R); accessing information via multiple modalities (C); converting data-centric language to user-friendly labels (A); and enabling switching between Olli and data table (F).

\section{\toolname{} System Interface \& Architecture}
\label{sec:system}
Below, we outline the input chart format for \toolname{}, explain how \toolname{} facilitates keyboard navigation and conversational interaction with the chart, and delve into further accessibility considerations integral to the design decisions outlined earlier.

\subsection{Input Chart Format}

\toolname{} operates on the premise that both the visual encoding information and the underlying dataset are accessible. In our work, we employ Vega-Lite specifications~\cite{satyanarayan2016vega} as the primary input for our system. Other chart specifications like Observable Plot~\cite{observableplot} and HighCharts~\cite{highcharts2022} can be adapted, provided they expose the underlying data and visual encoding variables. New adapters will need to be written to work with Olli, which currently supports Vega \& Vega-Lite, and Observable Plot~\cite{observableplot}. Alternatively, charts can be translated to the Vega-Lite specifications to be directly used in \toolname{}. The symbolic representations underlying the chart are parsed to create a keyboard-navigable tree view and to provide useful information about the chart's appearance for the question-and-answer pipeline, which is detailed below.

\subsection{Exploring Chart Encoding using Keyboard}
\label{sec:chart-encoding-navigation}
\begin{figure*}
  \includegraphics[width=\textwidth]{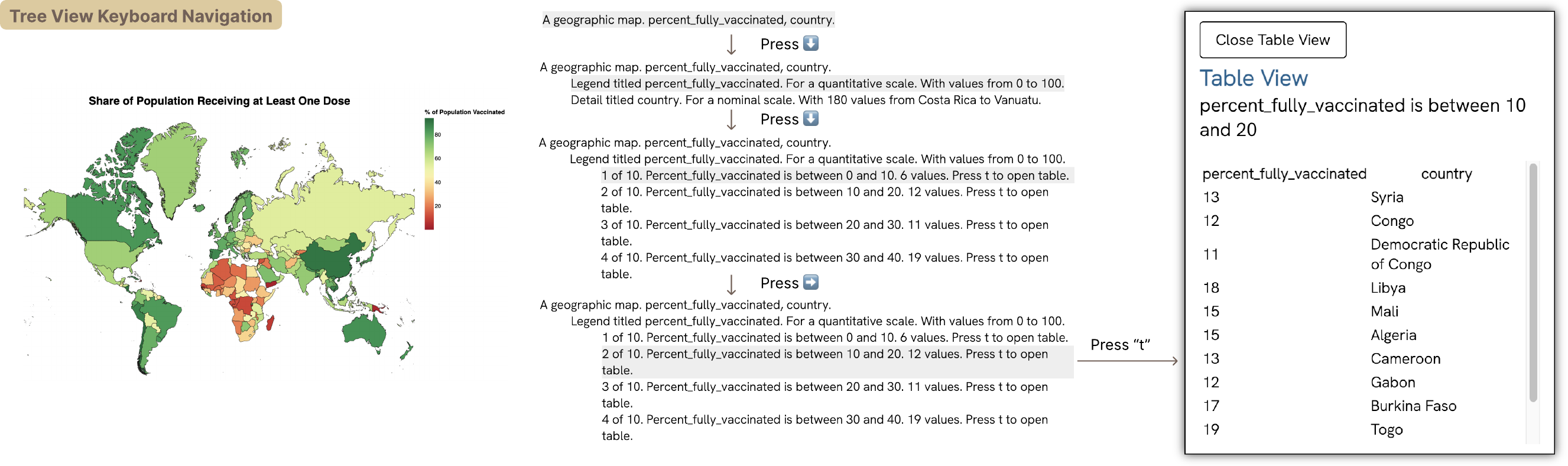}
  \caption{An example of a user's keyboard traversal of the Olli Tree. Users can widen/narrow the scope of the text via the up/down arrow keys (respectively), and otherwise navigate between sibling nodes using left/right arrow keys. To access individual data, users can press the 't' key to view a snapshot data table}
  \Description{A choropleth map is displayed on the left, with a guided traversal of the Olli Tree being shown to the right of it. Segmented into 3 steps, the guided traversal shows the necessary actions it takes to get from the header to the interval of percent fully vaccinated from 10 to 20. Off to the right is a table view of the data belonging to this interval.}
  \label{fig:olli}
\end{figure*}

We leverage Olli~\cite{2022-olli} to make the chart encoding explorable---\hlgold{D1}, as it provides an off-the-shelf, open-source solution based on Vega-lite. Olli renders a visual chart for sighted users and a keyboard-navigable tree view featuring chart descriptions at various levels of detail (see \autoref{fig:olli}). A more detailed explanation of Olli, including its design considerations, supported chart types, and empirical evaluation, is available in prior work~\cite{2022-olli,zong2022rich}.

Olli's tree view displays the chart content in a hierarchical structure, starting with the chart type description at the root, followed by visual encoding channels such as axes and legends. Within each encoding channel node, Olli lists data categories or numerical ranges depending on the data type being encoded; e.g., for a color legend, it lists all categories in the legend. Individual data points reside in these group nodes. All four chart types we used in this work, including line chart, bar chart, scatter plot, and choropleth map, had four levels of information granularity.

Based on its hierarchical structure, users can navigate the different levels of the tree view using up and down arrow keys (\textit{barchart} $\rightarrow$ \textit{legend} $\rightarrow$ \textit{negative polarity}) while using left and right arrow keys to navigate sibling nodes within each level (\textit{negative polarity} $\rightarrow$ \textit{positive polarity}). In order to access individual data points, Olli requires users to press \textit{t} to open up a screen-reader-compatible data table. This table shows a subset of the whole data, only displaying data points within the selected category or numerical range. 

The current version of Olli \textbf{does not support navigating a choropleth map by geographic regions}. We extended it to support the level of \textit{detail} channel in Vega-lite\footnote{\url{https://vega.github.io/vega-lite/docs/encoding.html\#detail}}. As a result, we can encode country names or state names into the \textit{detail} channel, which is in turn converted into an additional encoding channel node (see \autoref{fig:olli}).

\subsection{Exploring Underlying Data Table via Keyboard}
\label{sec:data-table}


\toolname{} offers users the flexibility to switch between the tree view and a conventional raw data table view (see options displayed in the buttons in \autoref{fig:teaser})---\hlgold{D5}. While the tree view facilitates structured exploration based on visual encoding, the data table provides additional advantages like sorting features, enabling users to quickly access specific data values and patterns. 
We disable navigation queries in the data table module as screen readers provide a slew of keyboard navigation shortcuts such as moving between headers and cells. The data table module supports the alphabetical/numeric sorting by each column.

\subsection{Rapid Chart Probing via Conversational Interaction}

\begin{figure}
  \includegraphics[width=0.95\linewidth]{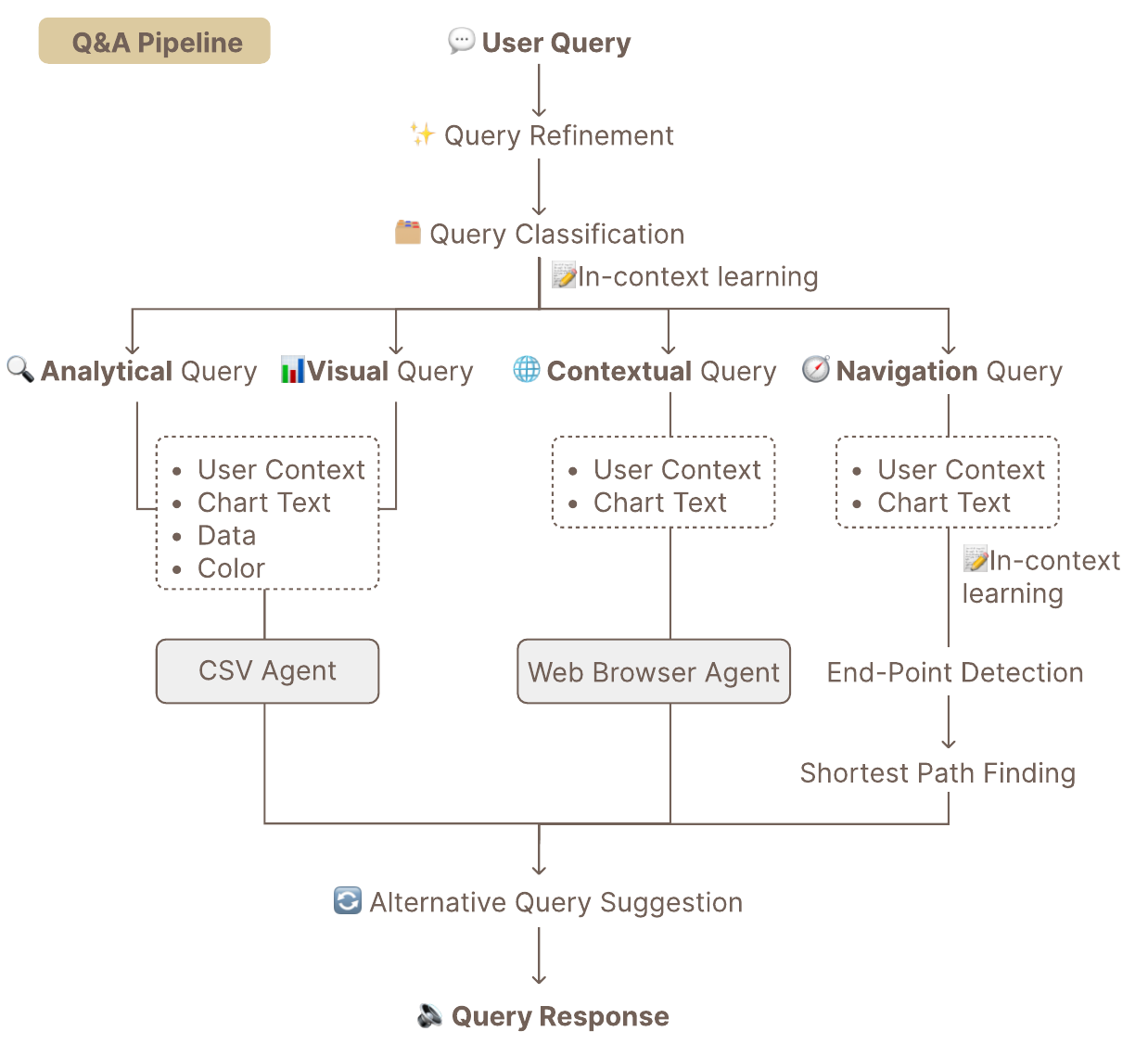}
  \caption{  \toolname{} pipeline takes a user query and refines it to improve clarity. The query is classified into one of four query types, each of which is fed to a different agent. If \toolname{} fails to respond, it attempts to suggest alternative queries.}
  \Description{The image illustrates a Q\&A Pipeline that processes user queries through refinement and classification, leading to analytical, visual, contextual, and navigation queries, and integrates in-context learning to generate responses. \nam{When did we detect unanswerable queries?}}
  \label{fig:pipeline}
\end{figure}

Keyboard navigation helps blind users grasp the chart's visual structure and data, but can be cumbersome for higher-level comprehension, such as computing aggregates or understanding overall visuals. We integrate LLMs for intuitive chart question-answering and enhance interaction by combining keyboard navigation with speech inputs. \autoref{fig:pipeline} shows the overview of the question-answering pipeline.






\begin{table*}[]
\centering
\begin{tabular}{lcccccl}
\toprule
Query Type              & Line Chart & Bar Chart & Scatterplot & Choropleth Map & Total \\
\midrule
\textsc{Analytical} Query & 147        & 168       & 242         & 193            & 750   \\
\textsc{Visual} Query     & 42         & 28        & 54          & 41             & 165   \\
\textsc{Contextual} Query & 16          & 14         & 28          & 12              & 70    \\
\textsc{Navigation} Query & 12         & 12        & 12          & 12             & 48    \\
\midrule
Total                     & 217        & 222       & 336         & 258            & 1033   \\
\bottomrule
\end{tabular}
\caption{\label{tbl-test-data} Distribution of data for four types of queries across different chart formats, including line chart, bar chart, scatterplot, and choropleth map, indicating the prevalence of analytical queries in the dataset.}
\end{table*}

\subsubsection{Data Set}

We utilized a prior study’s data set~\cite{kim2023exploring}, comprising 979 BLV user questions spanning four visual stimuli (bar, line, scatter, and map) for the development and quantitative evaluation of \toolname{} (\autoref{tbl-test-data}). We then partition the pool of questions once more into an 80/20 split between the testing and validation sets via stratified random sampling so that there is a proportionate representation of each query type amongst both sets. 

The questions were gathered through a wizard-of-oz study, where a human facilitator acted as a question-answering system. We reconstructed the visualization images into Vega-Lite specifications and partitioned the questions into \textsc{analytical}, \textsc{visual}, and \textsc{contextual} queries, which we derived from the question taxonomy in prior work~\cite{kim2020answering,kim2023exploring}. In addition, we employed GPT4 to generate 12 example \textsc{navigation} queries for each of the four visual stimuli, appending them to the prior data set of 979 BLV user questions. We define each query type in Section \ref{sec:supported-query-types}.

\paragraph{How did we derive the question taxonomy?} The original taxonomy by Kim et al.\cite{kim2023exploring} offers more fine-grained categories, such as granular analytical tasks like retrieving values and finding extremum\cite{amar2005}. They also categorized queries as visual vs. non-visual and look-up vs. compositional, similar to previous chart QA systems~\cite{kim2020answering}, while separating non-data queries. In our work, we reclassify these queries, as the detailed categorization is not needed for users and may not enhance the system's performance, considering the advanced language comprehension capabilities of LLMs.
First, we consolidated data-driven and visual tasks into \textsc{analytical} (or non-visual) and \textsc{visual} queries, respectively. We do not separately consider look-up vs. compositional, as we expect LLMs to handle them without differentiation. Finally, we consolidated non-data queries into \textsc{contextual} queries.

\paragraph{What additional metrics did we introduce?} We also elaborate on the previous taxonomy by introducing two new binary metrics: \textsc{open-endedness} and \textsc{answerability}. Open-ended queries allow for multiple valid interpretations and responses, and often invoke questions like ``Why?''. Open-ended questions may also involve computations, especially when the operation parameters are ambiguous: ``And what was that temperature?''. Answerable queries are relevant to the data set. By contrast, unanswerable queries are irrelevant and cannot be feasibly addressed using the corresponding chart. 

\paragraph{How did we generate navigation queries?} We used GPT-4 with few-shot prompting to generate  \textsc{navigation} queries with an equal distribution of orientation and wayfinding queries. 
To simulate real user chart interaction, we include the corresponding active element in the tree view for each query.

\paragraph{How did we generate ground-truths?} The ground truths for the testing and validation sets were manually generated by two researchers independently and then merged by resolving conflicts. 
The process involved reading charts, calculating numbers, and searching for information online. Throughout the generation process, we emphasized verboseness. For instance, the ground truth response to the question ``What is the vaccination rate of South Africa'' is ``The vaccination rate for South Africa is 36\%'', as opposed to the more concise ``36\%''. The ground truth responses for \textsc{navigation} queries consist solely of starting and ending nodes to prioritize the model's ability to translate ambiguous user wording into precise nodes within the tree view.


\subsubsection{Supported Query Types}
\label{sec:supported-query-types}

\textbf{\textsc{Analytical} queries}---\hlmaroon{D2} primarily focus on understanding the underlying data, such as ``Is Africa the country that needs the vaccine the most?'' or ``What is the highest positive anomaly?'' \textbf{\textsc{Visual} queries}---\hlmaroon{D2} relate to visual encoding information or demand visual interpretation, exemplified by questions like ``What color is North America?'' or ``Is the line fluctuating?'' \textsc{Analytical} and \textsc{visual} queries are not entirely distinct; \textsc{visual} queries often necessitate data interpretation, as in ``Which country exhibits the darkest shades for both the lowest and highest values?''. Despite this overlap, we retain a conceptual separation in order to communicate to LLMs that a query involves information
encoded visually and gain a more fine-grained understanding of VizAbility’s performance with chart-specific queries.



\textbf{\textsc{Contextual} questions}---\hlmaroon{D3} seek information not directly present on the chart but require ancillary knowledge related to it. For instance, some questions aim to understand the chart's encoding, like ``What is a scatterplot?'' or ``What does `positive temperature anomaly' mean?'' Others ask about context related to the data, such as ``Where is Palestine?'' or ``Why does the data start in 1880? What occurred then?'' Additionally, there are inquiries about the data's origin, exemplified by ``What is the source of this information?'' or ``From where was this data obtained?''

\textbf{\textsc{Navigation} queries}---\hlmaroon{D4} are a category we introduced to enhance user experience. These queries are tailored to the synergy between keyboard navigation and conversational interaction. For instance, to reduce cumbersome keyboard navigation and assist users in orientation, questions such as ``How can I get to the X-axis'' (wayfinding) or ``Where am I?'' (orientation) can be beneficial. Our motivation for this stems from a previous empirical study~\cite{kim2023beyond}, where blind individuals highlighted such challenges with Olli's tree view.

\subsubsection{Query Classification}
\begin{figure*}
  \includegraphics[width=\textwidth]{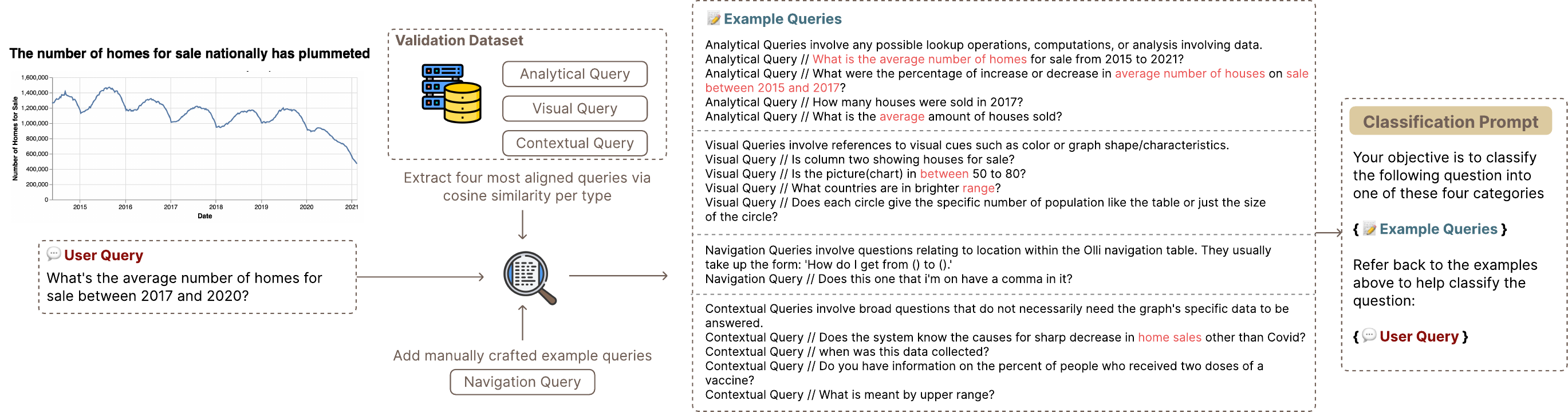}
  \caption{User questions are initially categorized based on query type via an LLM trained with few-shot prompting. We populate the prompt with sample questions and their corresponding ground truth classifications, which we extract from the validation set. Only those validation questions that share the highest cosine similarity score with the user query are selected within each query type.}
  \Description{The line chart is displayed off to the left. To the right of it lies an overview of the classification pipeline, beginning with the user query and validation data set, and ending with a few-shot prompt example off to the right of the figure.}
  \label{fig:classification}
\end{figure*}

First, we aim to classify user queries based on this categorization rather than diving straight into responses. Once classified, we proceed to address each type of query in the subsequent phase (see the next section). This task division provides the LLM with a well-defined task and has been proven to increase its performance~\cite{wei2022chain}, while also enabling more efficient allocation of computational resources for each type of query. \autoref{fig:classification} shows our few-shot prompting approach. In the prompt, we provide a clear definition for each query type. To bolster the definition, we accompany each with four exemplar questions. 

These examples are sourced from our validation set, chosen based on their close alignment with the user query at query time. Specifically, for each query type and the given user query, we sift through the validation set to pinpoint the four most analogous queries. These are then incorporated as representative examples for each query definition within the prompt. For this endeavor, we used sentence transformers~\cite{reimers2019sentence} to generate text embeddings and then applied cosine similarity to these embeddings to identify the most closely aligned examples. This method offers greater precision compared to arbitrarily selecting samples for each query type.

\change{
}

We constrain the range of LLM responses by explicitly instructing it to output either: ``Analytical Query'', ``Visual Query'', ``Contextual Query'', or ``Navigation Query''. To thwart any potential hallucinations from the LLM, we provide an accessible escape route by instructing the model to return ``I am sorry. I am unable to answer this question'' when confronted with a question that does not immediately conform to any of the specified query types.  Without such a safeguard, GPT frequently generates technical jargon and error messages that can deter users.

\subsubsection{Query-Specific Prompting}
The answering pipeline diverges into three unique paths, depending on the query type (\autoref{fig:teaser}).

\paragraph{\textsc{Analytical} \& \textsc{Visual} Queries}

\begin{figure*}
  \includegraphics[width=\textwidth]{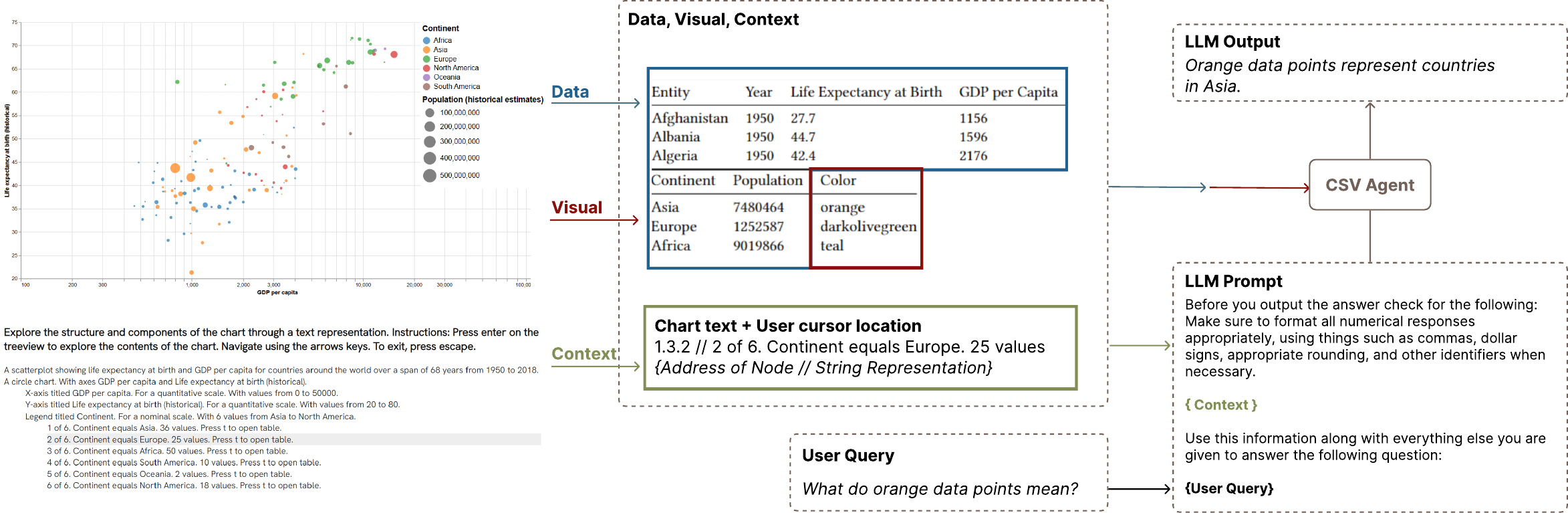}
  \caption{Query-specific evaluation for Analytical and Visual queries. We parse the chart's transformed data set and aggregate color encoding within a CSV file, which we then supply to an LLM via a CSV agent. For further context, we also populate the prompt with the user's active position within the Olli Tree, in addition to a text representation of the Tree itself.}
  \Description{An overview of the pipeline for evaluating Analytical/Visual queries. The figures is divided into three columns, with the leftmost column depicting a scatterplot and accompanying Olli Tree. The middle column shows the inputs sent to the LLM (transformed data, color encoding, and active user position), whereas the rightmost column shows the LLM sample prompt.}
  \label{fig:analytical-visual-query}
\end{figure*}

\hlmaroon{D2}---\autoref{fig:analytical-visual-query} illustrates our approach to handling \textsc{analytical} and \textsc{visual} queries. To circumvent the predefined token limit of the LLM, we consolidate the transformed data extracted from the Vega View~\cite{vegaview} into an external CSV file. This file is then processed by LangChain's CSV Agent~\cite{csvagent}, which operates in the background. Under the hood, this agent leverages the Pandas DataFrame agent, subsequently executing Python code generated by the LLM. We purposefully avoid including the entire raw dataset, recognizing that it might differ from the final view data. Often, the agent can get stuck in an infinite loop of thinking. To prevent this issue, we have implemented a time constraint. If the time limit is exceeded, \toolname{} will display the message, ``Answer: I'm sorry, but the process has been terminated because it took too long to arrive at an answer,'' and will also suggest alternative questions to the user (see Section \ref{sec:mitigation}).

While the CSV agent can handle most data-related queries, it is not aware of any visual encoding information of the chart. To address visual queries, we extract color information directly from the Vega View~\cite{vegaview} and incorporate it as an additional column within the CSV file. This modification ensures that each data point is paired with its corresponding color. Initially, the extracted color data is in hex codes. To enhance user-friendliness, we employ a color-matching algorithm to convert the hex codes into more common English names. This algorithm works by cross-referencing the source hex code with a predefined list of color hex codes and English names~\cite{csscolors}, ultimately determining the closest matching name based on their relatives distances within the CIELAB color space.


The color augmentation process enables answering visual questions like ``What color is Algeria? What other countries are the color of Algeria?'', as \toolname{} responds: ``Algeria is orange-red and other countries with the same color are Syria, Iraq, Congo, [...].'' Furthermore, LLM is lenient with user queries and accepts a certain margin of error for color input. e.g., if the user asks about what \textit{blue} represents, the system can infer \textit{blue} refers to \textit{steelblue} in the map.


To provide further visual context for the chart, we have integrated a textual representation of the chart generated by Olli directly into the LLM prompt (see \autoref{fig:analytical-visual-query}). This addition has the potential to significantly enhance the performance of visual question-answering. For example, when presented with the question ``What does the graph show?'', the system, without the text representation, provided a response like ``The graph shows the data from the dataframe, which includes the year, value, temporal polarity, ...''. However, when furnished with the text representation, the LLM responded with a more comprehensive and human-friendly answer: ``The graph shows the temporal polarity of the temperature anomaly (in degrees Celsius) from 1850 to 2021 and the y-axis representing the temporal anomaly in degree Celsius. [...]''


Moreover, we supplement it with the user's current position within the tree view, tracked via the user's keyboard movements. This feature can help address potentially ambiguous questions. For instance, a user might ask, ``What's an average?'' with the intention of inquiring about the average within a category where their cursor is located. We also ensure that the responses are properly formatted with commas and special characters so that they are optimized for screen reader interpretation. For example, we present the number 468297 as 468,297 to improve clarity, especially since NVDA, the most popular screen reader~\cite{webaim2023screen}, would otherwise read it as 'four six eight two nine seven' in its default settings, which could be less intuitive for users.

\paragraph{\textsc{Contextual} Queries} \hlmaroon{D3}---To address contextual queries that require background or external information on what is available in the chart or its data, we have incorporated a Web Browser agent~\cite{serpapi} to retrieve more general information relevant to chart comprehension. 
For example, when presented with the contextual query, ``What do you mean by temperature anomalies,'' the LLM responds with, ``Temperature anomalies are any measure of temperatures that are unusual for a particular region, season, or time period. [...]'' Categorizing questions beforehand enabled us to streamline the process and eliminate unnecessary, resource-intensive prompts needed for analytical and visual queries. 

Often, contextual queries require information about the chart or data to disambiguate user queries. For example, queries such as ``Does the system know the causes for the sharp decrease in home sales other than Covid?'' or ``When was this data collected?'' need to understand what the chart is about, without requiring detailed underlying data. Therefore, we accommodate these instances by incorporating the readily available high-level text representation of the chart into the Web Browser Agent.

\paragraph{\textsc{Navigation} Queries}
\label{sec:navigation-queries}
\begin{figure*}
  \includegraphics[width=\textwidth]{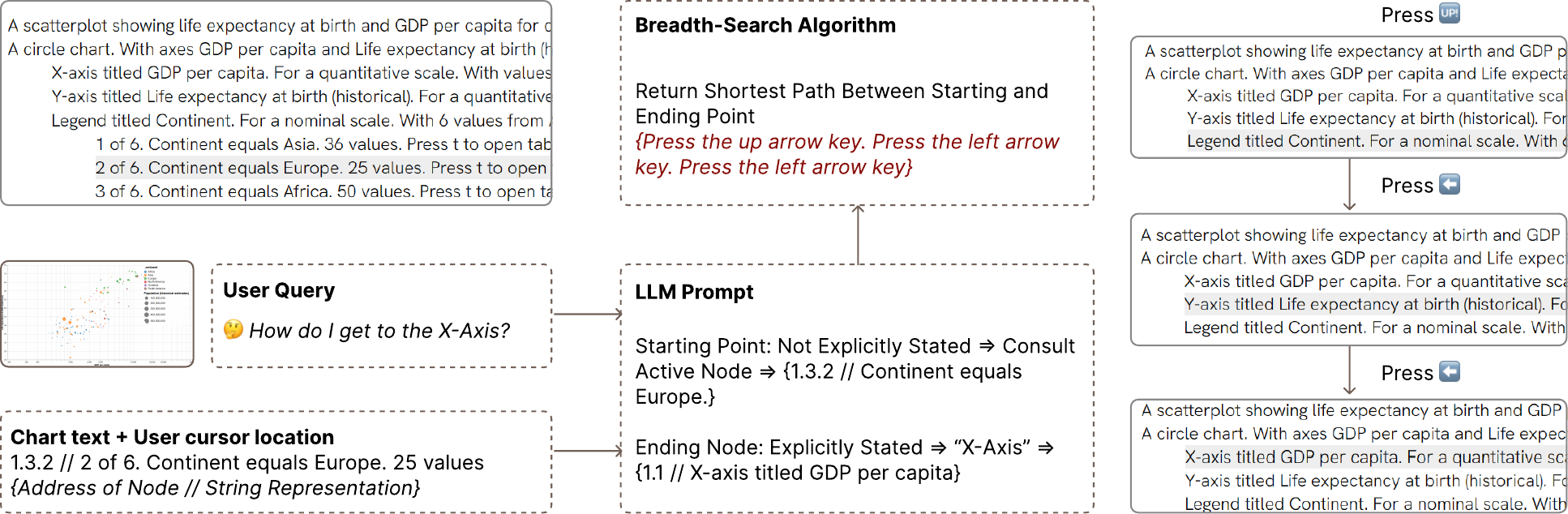}
  \caption{Query-specific evaluation for Navigation queries. We pass a text representation of the Olli Tree and the addresses of corresponding nodes within the Tree to an LLM alongside the user question. With the aid of few-shot prompting, the LLM then identifies the starting and ending nodes within the Olli Tree. Should the starting node not be explicitly mentioned within the question, the model instead utilizes the user's current location within the Tree. We then execute a breadth-search algorithm and relay the shortest path between starting and ending nodes back to the user.}
  \Description{An overview of the pipeline for navigation query evaluation. The figure is divided into three columns, with the leftmost column displaying the user query and Olli Tree, the middle column showing a sample LLM prompt, and the rightmost column showing the traversal of the Olli Tree (following the instructions outputted by the LLM)}
  \label{fig:analytical-visual-query}
\end{figure*}

\hlmaroon{D4}---We seek to integrate users’ keyboard navigation with the conversational module via navigation queries.
\toolname{} currently supports two types of navigation queries: (a) \textbf{wayfinding} questions, in which, upon being provided a starting and ending point within the tree view, the model returns a series of directions dictating the shortest traversal and (b) \textbf{orientation} questions, in which the \toolname{} returns the user’s current location within the tree view. 

To handle navigation queries, we attribute a unique address to each node of the tree view and convey this address, along with the user’s current position, to the LLM. Through the utilization of few-shot prompting, we instruct the LLM to discern the starting point and ending point from the user query. It is important that the model has significant leniency in user queries, as it is highly unlikely that the user will specify the exact starting/ending points verbatim. Thus, the few-shot prompting primes the LLM to properly interpret the user query. For example, in response to the query ``Take me to Haiti'' (related to the choropleth map), the LLM comprehends the user query's context and correctly deduces that the absence of an explicit starting node means the user intends to initiate navigation from their current location. On the other hand, \toolname{} can easily infer the ending point, which is the node titled: ``3 of 180. Country equals Haiti. 1 value. Press t to open table.'' If the model cannot discern any starting or ending point, it yields: ``The question was interpreted as involving navigation, but either no starting/ending point was provided, or the tree view was not activated. Please try again.''

Once the starting and ending points have been identified, we employ a breadth-search algorithm that returns string instructions of the shortest path, which users can then manually follow at their own discretion. To avoid repetition we coalesce instructions whenever possible. For instance, for instructions pertaining to the navigation between two sibling nodes, we convert ``Press the right arrow key. Press the right arrow key. Press the right arrow key.'' to ``Press the right arrow key 3 times.'' We initially opted for this approach as opposed to automatically moving the user to their desired ending point with the rationale that autonomy and transparency are crucial for our intended audience.


\subsection{Mitigating LLM Failures through Responsive Feedback}
\label{sec:mitigation}

Several factors can create a disconnect between the user's query and the output from the LLM. These factors might be technical, such as when the CSV Agent executor is prematurely terminated for exceeding its time limit, or they might arise from the ambiguity of the user's query and the system's consequent struggle to categorize and address it. To tackle these challenges, we implement strategies that are both proactive (occurring before classification and answering) and reactive (occurring after these processes), aiming to create a more user-friendly environment.

\subsubsection{Proactive Error Mitigation Strategies}
\hlmaroon{D6}---The versatility of \toolname{} poses novel challenges in user query interpretation and answering. As reflected in the validation data set, some user queries may be too broad, ambiguous, or can even refer to variables that are not explicitly present in the data. For instance, the real user query ``What kind of vaccine they used?'' is irrelevant to the choropleth map displaying the share of the population that received at least one dose of the Covid-19 vaccine. User queries can also be poorly worded, such as ``What parts of North America are not in the 80 to 100 percent range'', deeming it more difficult for an LLM to accurately interpret and compute the answer. These irrelevant and ambiguous queries can significantly decrease the overall accuracy of the system if not accounted for.

To mitigate the effects that ambiguous queries may have on the system's overall accuracy, we introduce an additional \textbf{query refining process} that occurs before user query classification. We furnish a GPT-3.5-Turbo prompt template with the user query and---serving as the primary source of context---a text representation of the actively engaged chart. The LLM is instructed to add to---but not alter---the question so that it is as specific and relevant to the data as possible, whilst still retaining its original meaning. Referring back to the earlier example of an ambiguous query, the LLM refines it to: ``Which countries in North America have a percentage of fully vaccinated individuals below 80\%''. The pipeline continues as detailed in prior sections, but now utilizes the refined query instead of the raw user query. Whereas before \toolname{} could not accurately answer the above question and instead outputted a list of all of the countries represented in the data, the system is now able to consistently identify all countries that occupy the desired range.

\subsubsection{Reactive Error Mitigation Strategies}
\hlmaroon{D6}---LangChain's CSV Agent leverages chain of thought reasoning administered over a series of sequential prompts (in conjunction with pandas dataframe) to answer questions pertaining to a CSV file. Each individual prompt must therefore follow a specified format of Observation, Action, and Action Input for a subsequent prompt to properly parse the result~\cite{csvagent}. We have observed that sometimes an individual prompt in this sequential chain may be outputted in an incorrect format, thus triggering a cascading effect that ultimately results in an OutputParserException. To mitigate this error, we utilize prompt engineering and direct the CSV Agent Executor ``You must follow the structure of Observation, Thought, Action, Action Input, etc.'' to ensure that the output format is consistent between sequential queries. If the CSV Agent Executor is active for an extended period of time, we manually terminate it to preserve time-efficiency. 

We handle these two potential errors by being explicit in our language; e.g., ``I am sorry but I cannot answer the question''. To foster an iterative and exploratory environment, we also employ a pipeline that recommends two questions that retain the essence of the original user query but eliminate error-inducing language. We achieve this recommendation by monitoring the response of the answering pipeline. If the output exhibits any sign of error (refer back to the above cases) or is otherwise incomplete, i.e. ``No answer can be found'', we populate a prompt template with the original user query, the error output (which identifies the error), and a text representation of the chart for additional context. This prompt is then passed through the LLM, which generates two new questions that are subsequently relayed back to the user in the form of clickable buttons.
For instance, the error-inducing question: ``Where are these houses sold?'' yields the following LLM output, ``The data does not contain any information about the location of the houses.'' and is accompanied by two rephrased queries: ``What information regarding the sale of these homes is provided in the dataset beyond the date and inventory quantities?'' and ``Could you elaborate on any additional details related to the properties or their characteristics available within the dataset?'', each displayed as a button.
\subsection{Fostering a More Accessible User Experience}
\label{sec:user-experience}

\paragraph{Providing different query methods and audible cues} \hlmaroon{D7}---Users can submit conversational queries via voice recordings that are processed via the Whisper speech recognition~\cite{whisper}. However, oftentimes, enabling microphones can be problematic. Thus, we provide an alternative text box so that they can type the queries using the keyboard. Upon inputting their question (regardless of the modality), users are provided with an audible cue of ``Loading. Please Wait''. Every subsequent 3 seconds, the user is exposed to yet another audible cue, this time ``Still Loading''. This loading cue significantly improves transparency and mitigates any possible confusion that can arise from an unresponsive webpage. 

\paragraph{Enhancing user trust and transparency in responses} \hlmaroon{D7}---\toolname{} does not solely display the answer, and instead provides the user query and brief justification behind its response in conjunction with the actual answer. For instance, the following is articulated by \toolname{} when a user asks, ``What is a choropleth map?'': ``Your question 'What is a choropleth map?' was categorized as being context-seeking, and as such, has been answered based on information found on the web.'' By letting users know the scope of the answer (i.e., whether it was sourced from the internet, data, or the tree view), we allow users to verify and evaluate the effectiveness of the LLM response independently, thus bolstering user trust and system transparency.

\paragraph{Offering cold-start query suggestions for onboarding}
\hlmaroon{D7}---To help users figure out what queries are possible, \toolname{} generates four initial queries, each of which belongs to one of the four query types. We achieve this suggestion by querying the LLM with a prompt that uses in-context impersonation~\cite{salewski2023context} (``Pretend you are a blind/low vision user who is presented with a chart.''), coupled with a text representation of the current chart. For instance, the LLM generates the following questions for the bar chart: ``What is the temperature anomaly for the year 2020?''; ``Can you provide a description of the color scheme used in the bar chart to represent the temperature anomalies?''; ``Are there any patterns or relationships between the year and the temporal polarity of the temperature anomaly?''; ``How do I get from my current position in the text representation to the x-axis?'' These questions appear as interactive buttons, allowing users to either choose a suggested question or input their own.

\subsection{Notes on Interactive Charts}
\toolname{} extends its capabilities to interactive charts, allowing for dynamic updates in both the tree view and question-answer components when users modify the chart. An example of this update is seen in the scatter plot (referenced in \autoref{fig:analytical-visual-query}), which includes a slider for filtering data by year. As a user selects a specific year, \toolname{} generates a new view of the data for the LLM and simultaneously updates the tree view to reflect the chosen year.

\section{Evaluation: Q\&A Performance Benchmark}




For our quantitative evaluation, we concentrated on validating the question-answering pipeline using the testing dataset. This evaluation comprised two components: assessing the accuracy of query classification and evaluating the correctness of question responses.




\subsection{Classification Evaluation}


Our evaluation yielded an overall classification accuracy of 87.39\%. \autoref{tbl:classification-result} presents detailed results, including precision, recall, and F1-scores for each class. \autoref{tbl:misclassification-examples} provides examples of misclassified instances.

\begin{table}[]
\centering
\begin{tabular}{@{}l|llll@{}}
\toprule
Query Type & Precision & Recall  & F1      & Count \\ \midrule
Analytical & 90.96\%   & 93.10\% & 92.02\% & 551   \\
Contextual & 64.65\%   & 67.37\% & 65.98\% & 95    \\
Navigation & 100\%     & 97.50\% & 98.73\% & 40    \\
Visual     & 89.09\%   & 74.81\% & 81.33\% & 131   \\ \bottomrule
\end{tabular}
\caption{\label{tbl-participants} Quantitative results contextualize \toolname{}'s classification accuracy of 87.39\% through the lenses of Precision, Recall, and F1 scores. The system is most proficient at classifying analytical and navigation queries, as indicated by the significantly higher F1 Scores attributed to these query types.}
\label{tbl:classification-result}
\end{table}

\begin{table*}[]
\centering
\label{tbl:misclassification-examples}
\begin{tabular}{lll}
\hline
True Class & Predicted Class & Example Text                                                                                                    \\ \hline
Analytical & Contextual      & Does the data include booster shot?                                                                             \\
Analytical & Visual          & Is Asia in the medium upper range in 60 to 80 area?                                                             \\
Contextual & Analytical      & What is the percentage at which vaccinated population reach herd immunity?                                      \\
Contextual & Analytical      & What is the source of this data?                                                                                \\
Contextual & Visual          & Describe a scatterplot.                                                                                         \\
Contextual & Visual          & Can I get a map of the US?                                                                                      \\
Visual     & Analytical      & How many years in total are represented on the X axis?                                                          \\
Visual     & Analytical      & Which part of Asia has a darker shade of green...                                                               \\
Visual     & Analytical      & What is the trend for Brown?                                                                                    \\
Visual     & Contextual      & Is there anything specific you want me to look for in this chart?                                               \\
Navigation & Visual          & ...am I closer to the start or the end of the Y-axis? \\ \hline
\end{tabular}
\caption{Additional Examples of Misclassification Cases}
\label{tbl:misclassification-examples}
\end{table*}

The model was effective in identifying analytical queries, with a 93.10\% ($\frac{513}{550}$) recall and 90.96\% ($\frac{513}{564}$) precision.  Analytical queries often contain distinct computationally-heavy language, such as ``correlation'' and ``average decrease,'' as seen in examples like ``What's the correlation between GDP per capita and population?'' and ``For the time from September 2019 to September 2020, what was the average decrease in homes for sale?''  This specific language may aid in differentiating these queries from broader contextual ones, potentially leading to a higher overall accuracy.

Of the $95$ queries designated as contextual in the ground truth, 67.37\% (64) were classified as such by \toolname{}. In addition to the 64 true positives, the model incorrectly classified 30 analytical queries and 5 visual queries as contextual in nature - contributing to a precision of 64.65\% ($\frac{64}{99}$). Poorly classified contextual queries like ``What is the source of this data?'' and ``Can I get a map of the US?'' may indicate that the model often emphasized individual words over the overall meaning of queries. For instance, the word ``data'' likely led the model to classify the first query as analytical. Similarly, the word ``map'' likely prompted the model to mistakenly identify the second query as visual.

A recall of 74.81\% or $\frac{98}{131}$ for visual queries indicates that \toolname{} correctly classified around $\frac{3}{4}$ of the queries designated as visual by the ground truth. On the contrary, visual queries such as ``Which part of Asia has a darker shade of green, which has the most vaccinated amount of people?'' and ``What is the trend for Brown?'' were falsely identified as being analytical, despite the presence of visual language $\{$``Brown'', ``darker shade'', ``green''$\}$. This lower performance may be due to the significant overlap between visual and analytical queries. For example, we consider any question that references visual components of the data (``Brown'') as visual, even if it also involves computations (``trend''). On the other hand, the model rarely misclassified analytical or contextual queries as visual, as demonstrated by an 89.91\% ($\frac{98}{109}$) precision score.

\toolname{}'s proficiency in distinguishing navigation queries may be attributed to the distinct and uniform structure to which most GPT4-generated example queries conform; i.e., queries either assume the role of wayfinding or orientation questions. The results, showing a 100\% ($\frac{39}{39}$) precision and 97.50\% ($\frac{39}{40}$) recall, suggest that \toolname{} is effective in distinguishing these tasks from the typical data retrieval or lookup associated with analytical and visual queries. Nonetheless, the misclassified query ``While exploring inventory values between 800000 and 1000000, am I closer to the start or the end of the Y-axis?'' hints that the model might sometimes give undue weight to individual words. For example, the reference to `the Y-axis', a visual element of the chart, could have led the model to categorize the navigation query as visual.



\begin{figure}
  \includegraphics[width=0.95\linewidth]{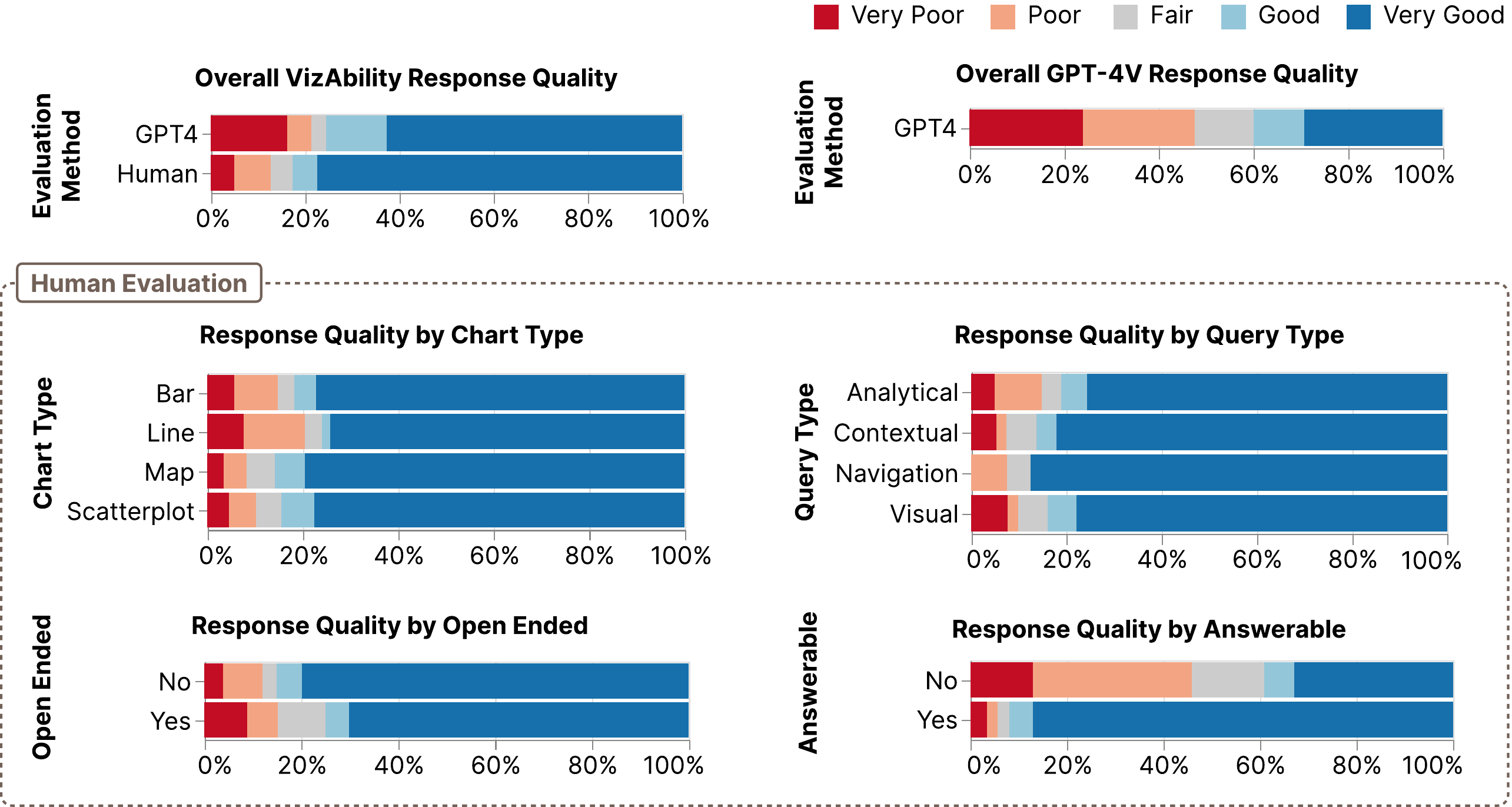}
  \caption{Quantitative results display the distributions of quality ratings (via a 5-point likert scale) for \toolname{} responses. For more granularity, the results are also partitioned by query type, chart type, and question characteristics. \toolname{}'s performance is considerably higher than the GPT-4V baseline.}
  \Description{The image contains four horizontal bar graphs comparing classification accuracy and response quality across chart types, query types, and levels of leniency, with varying proportions of correctness and quality ratings from very poor to very good.}
  \label{fig:quantitative-result}
\end{figure}

\subsection{Question Response Evaluation}
Our evaluation of the response quality is twofold. First, two researchers manually inspected the quality of the responses. Based on this ground-truth evaluation, we established an LLM-based evaluation using GPT-4, which is intended to facilitate scalable, automatic evaluation for efficient benchmarking tests, following recent literature in natural language evaluation~\cite{liu2023gpteval,wang2023chatgpt,fu2023gptscore,li2024leveraging}. We describe each of these procedures below.

\subsubsection{Human Evaluation}

We evaluated each pair of system response and corresponding ground truth individually based on a five-point Likert scale: [Very Poor, Poor, Fair, Good, Very Good]. Our focus was on a response's `correctness' in terms of its coherence and consistency with the ground truth. Our assessment scheme favored explanatory responses over overly brief ones, in line with our recognition of trustworthiness and transparency as essential factors. \autoref{tbl-delineation} shows how we defined each increment on the Likert Scale, which is further elaborated in the GPT4-based assessment in Section \ref{sec:gpt4-eval}.


\begin{table*}[]
\begin{tabular}{l|l}
\toprule
Very Good & Response B is not only similar but also faithful to Response A.                                                                                                                          \\ \hline
Good      & \begin{tabular}[c]{@{}l@{}}Response B is mostly similar to Response A \\ but may lack some of the more specific key details such as labels.\end{tabular}                                 \\ \hline
Fair      & \begin{tabular}[c]{@{}l@{}}It is unclear whether there are any similarities between Response A and \\ Response B due to the ambiguity of one or two of the Responses.\end{tabular}       \\ \hline
Poor      & \begin{tabular}[c]{@{}l@{}}The content of Response B is somewhat irrelevant to that of Response A; \\ the core information in Response B does not match that of Response A.\end{tabular} \\ \hline
Very Poor & The content of Response B is irrelevant to that of Response A and bears no similarities.                    \\               \bottomrule                                                             
\end{tabular}
\caption{\label{tbl-delineation}  A delineation of the gradations used in the Likert scale to assess the degree of coherence between Response A and Response B, ranging from 'Very Good' to 'Very Poor'.}
\end{table*}

Of the $817$ user queries, $632$ or $77.36\%$ were deemed ``Very Good'' in the human assessment. Responses rated as ``Very Good'' often restated the user's question, formatted quantitative data correctly, and included contextual information. For example, in response to the query ``What continent has the highest vaccination rate?'' related to a choropleth map, \toolname{} answered, ``The continent with the highest vaccination rate based on the percentage of fully vaccinated people in each country is South America, with an average percentage of 69.7\%.'' This response, more detailed than the ground truth ``The continent with the highest vaccination rate is: South America'', demonstrates \toolname{}'s ability to provide comprehensive answers. The distribution for Good, Fair, and Poor responses was $5.14\%$ or $\frac{42}{817}$, $4.65\%$ or $\frac{38}{817}$, and $7.71\%$ or $\frac{63}{817}$ respectively. The human assessment yielded $42$ or $5.14\%$ of questions as being ''Very Poor'' in coherence to the ground truth. We elaborate on these findings and investigate relationships between the types of user queries and \toolname{}' responses below (\autoref{fig:quantitative-result}).

\paragraph{Partitioning based on query classification} \toolname{} exhibited the greatest accuracy in answering navigation questions. 87.5\% ($\frac{35}{40}$) of navigation queries received a ``Very Good'' assessment, which equates to the correct translation of the user query into concrete starting and end nodes within the tree view. For instance, \toolname{} correctly parsed the question ``What's the quickest path to get from the top of the tree to inventory values above 1400000?'' and identified ``Starting Point: `A line chart. With axes Date and Number of Homes for Sale'; Starting Address: 1; Ending Point: `Inventory is between 1400000 and 1600000'; Ending Address: 1.2.6'', which corresponds with the ground truth. User references to the tree view may not always be explicit (``top of the tree'' $\rightarrow$ ``a line chart...''). 82.11\% ($\frac{78}{95}$) of responses to contextual queries were identified as being ``Very Good''. This metric was slightly lower for both visual and analytical queries, from which 77.86\% ($\frac{102}{131}$) and 75.68\% ($\frac{417}{551}$) of responses had warranted a ``Very Good'' assessment respectively. Nevertheless, the distribution of assessment scores is relatively consistent across query types, with a tendency to skew towards ``Very Good'' (\autoref{fig:quantitative-result}).


\paragraph{Partitioning based on chart types} The distribution of assessment scores is similar across chart stimuli, suggesting that chart type may not have as significant an influence on overall system performance compared to other factors. Questions pertaining to the choropleth map had the highest frequency of yielding ``Very Good'' responses (79.61\% or $\frac{164}{206}$). This is followed by responses for the scatter plot (77.65\% or $\frac{205}{264}$), bar graph (77.27\% or $\frac{136}{176}$), and line chart (74.27\% or $\frac{127}{171}$). The low variability amongst chart stimuli highlights \toolname{}'s versatility in addressing a wide range of data visualizations.

\paragraph{Partitioning on the \textsc{answerability} of queries} We found that of all the questions for which \toolname{}'s response was prescribed a ``Very Poor'' rating, almost half ($45.24\%$ or $\frac{19}{42}$) had also been identified as being unanswerable (\autoref{fig:quantitative-result}). More notably, 76\% ($\frac{48}{63}$) of ``Poor'' responses also happened to be unanswerable. For instance, referring to the choropleth map which depicts the percentage of the global population receiving at least one dose of the vaccine, the query ``Which country has the fewest vaccinations?'' is unanswerable because of the absence of precise counts of vaccines from the data set. Nonetheless, \toolname{} fails to acknowledge that relative percentages do not equate to precise counts and generates the ``Very Poor'' output ``Among the listed countries, Burundi has the fewest vaccinations.''. Instances like these skew the overall accuracy of the system towards ``Very Poor''. By isolating just the answerable queries and recomputing accuracies, we find that 87.03\% ($\frac{584}{671}$) of responses are now ``Very Good'' and consistent with the ground truth, increasing the mean assessment score from $4.42$ to $4.70$.

\paragraph{Partitioning based on the \textsc{open-endedness} of queries} 70.10\% ($\frac{143}{204}$) of system responses to open-ended queries were prescribed a score of ``Very Good''. For the assessment of open-ended contextual queries, we emphasized \toolname{}'s ability to not only draw on external information, but also support claims with relevant data derived from the chart. The contextual query, ``Is there any other places that I can get some new information like a graph which shows by each region of the US?'' with ground truth: ``Yes, charts depicting inventory of houses for sale by region can be found on the internet.'', received the following ``Very Good'' response: 

\begin{quote}``Yes, you can find a similar graph that displays home sales data by different regions of the US. The existing dataset provides a line chart showing the number of homes for sale over time, and this type of data is often collected and analyzed at both the national and regional levels. \textit{Organizations such as the National Association of Realtors, Zillow, and Redfin} regularly publish reports [...]''
\end{quote}

In this response, \toolname{} not only explicitly identifies credible sources that the user can consult, but also clarifies those sources' relevance to the line chart. Nonetheless, \toolname{} proved to be more proficient in answering questions that were not open-ended. Of the $613$ user questions that are narrower in scope (not open-ended), $79.77\%$ yielded ``Very Good'' system responses - a value almost 10\% higher than the reported accuracy for open-ended queries (\autoref{fig:quantitative-result}). This performance increase may be due to the narrow scope of non-open-ended questions, which could reduce the chances of computational or logical errors; e.g., scatter plot questions such as ``What color is North America?'' are straightforward.
Consequently, \toolname{}'s response, ``The color that represents North America in the dataset is red.'' is almost verbatim to the ground truth, ``The color of North America on the scatter plot is: Red''. Similarly, responses to non-open-ended queries were also less frequently assessed as ``Very Poor''; 3.92\% vs. the rate of 8.82\% for open-ended queries.


\subsubsection{Automatic Evaluation using GPT4}
\label{sec:gpt4-eval}
We leverage LLMs for holistic evaluations that align better with human judgments compared to traditional metrics like BLEU, ROUGE, and BERTScore, which focus on simple text-level differences~\cite{li2024leveraging}. 

Our evaluation prompt, inspired by Liu et al.~\cite{liu2023gpteval}, presented two responses to GPT4 (gpt-4-0125-preview): Response A and Response B, with Response A acting as the ground truth. GPT4 was directed to assess the coherence of Response B in relation to Response A. We refrained from revealing which response was the ground truth or our own creation, as we hypothesized this indication to be extraneous information for the LLM, given its task to solely assess the coherence of two responses. The lower Kendall $\tau$ score ($=0.2900$) with human ground-truths for a similar prompt during our iterative testing, which explicitly defined Response A as the ground truth, supported this decision. 

Given that the coherence metric serves as an umbrella term for response evaluation, encompassing related aspects such as correctness, phrasing, and verbosity, we further prime the LLM through few-shot prompting with references. We populate the prompt with a manually selected sample of questions, \toolname{} responses, and their corresponding ground truth that we derive from the validation set, along with example scorings. To further contextualize the LLM's evaluation responses, we direct it to append a one-sentence rationale to its Likert score. During prompting, if the score deviated from the 1-5 range, GPT4 reassessed its evaluation of Response B. The results were formatted as \texttt{Score: $\{$coherence score$\}$ Rationale: $\{$rationale behind coherence score$\}$}.

Our automatic evaluation prompt yielded a Kendall's $\tau$ score of 0.5526 ($p<0.001$), signifying a strong correlation ($|\tau| \in [0.3,1.0]$) between the automatic assessments and human evaluation~\cite{chiang2023can}. The exact distribution of score ratings for GPT4 assessments is as follows: ``Very Poor'': 16.52\% or $\frac{132}{817}$, ``Poor'': 5.02\% or $\frac{41}{817}$, ``Fair'': 3.18\% or $\frac{26}{817}$, ``Good'': 12.85\% or $\frac{105}{817}$, ``Very Good'': 62.42\% or $\frac{510}{817}$. GPT4 demonstrated more variability in its assessments, with fewer outputs clustered around the ``Very Good'' rating. More precisely, we observed a reduction of 122 ``Very Good'' assessments and an increase of 63 ``Good'' assessments between the human and LLM evaluations, which might signify that GPT4 had adopted stricter and more rigourous criteria. For the choropleth map query, ``What is the highest vaccination rate in Africa?'', which yielded the system response, ``The country with the highest vaccination rate in Africa based on the geographic map dataset is Rwanda.'', the human evaluation designated this response as ``Very Good'' in overall quality due to its clarity, explanatory language, and close correspondence with the ground truth: ``The highest vaccination rate in Africa is 78.00\% and belongs to Rwanda.''. Nonetheless, GPT4 attributed a score of ``Good'', citing ``Both Response A and Response B identify Rwanda as the country with the highest vaccination rate in Africa, although Response B does not provide the specific percentage rate.''

\subsection{Baseline Comparisons}

\subsubsection{Comparison to a question-answering system with symbolic inputs} We compared our system to a similar system that focuses on chart reasoning~\cite{kim2020answering}. Directly running the system proved challenging due to compatibility issues with outdated dependencies. A similar evaluation, using our dataset from blind participants, was conducted in prior work, revealing an overall factual accuracy rate of 16\%~\cite{kim2023exploring}. This evaluation, however, was limited to just 245 queries from the total dataset. It excluded \textsc{contextual} queries and others deemed unanswerable due to ambiguous wording. Furthermore, the evaluation focused solely on questions related to bar and line charts, aligning with the system's supported question types. 

Overall, the system demonstrated some level of proficiency in answering straightforward value retrieval and extrema questions. The relatively low performance was mainly attributed to query comprehension and handling of diverse task types (e.g., yes/no and range questions). Seeking to maintain consistency with the prior system, we extracted data solely from the bar and line charts for a more fitting comparison. When narrowing the scope to these two types of visual stimuli, \toolname{} reports $\approx 76.01\%$ accuracy for the line chart and $\approx 81.82\%$ for the bar chart, only considering `Very Good' and `Good' responses, signifying a significant improvement in user query handling. 


\subsubsection{Comparison to GPT-4V(ision) with image inputs} We conducted a comparative analysis between \toolname{} and GPT-4 with Vision. Supplying GPT-4V with input images of the line chart, bar graph, scatter plot, and choropleth map, we generated responses to all $777$ analytical, contextual, and visual queries in the test set. We assessed the quality of GPT-4V responses using our GPT4-based automatic evaluation pipeline, which has been verified to correlate strongly with human evaluations. We omitted navigation queries because it is infeasible to provide just one reference image for the keyboard-navigable tree view. On the other hand, supplying the tree view text is essentially equivalent to using VizAbility. 

The GPT4 automatic evaluation yielded the following distribution of scores: 29.86\% or $\frac{232}{777}$ of responses received a ``Very Good'' assessment. 10.55\% or $\frac{82}{777}$ of responses were ``Good'', whereas 12.23\% or $\frac{95}{777}$ were ``Fair''. ``Poor'' and ``Very Poor'' responses occurred at equal frequencies: 23.68\% or $\frac{184}{777}$. This discrepancy in performance between GPT-4V and \toolname{}, which outputs ``Very Good'' responses at a frequency of 77.36\%,  can be attributed to GPT-4V lacking access to the raw data required for queries that require symbolic computations, resulting in inconclusive responses for certain prompts, such as ``What is the average vaccination rate for countries in the world?'':

\begin{quote}``The image you've provided is a world map that shows the share of the population receiving at least one dose of a COVID-19 vaccine... To determine the average vaccination rate for countries around the world, one would need to access the raw data from a reliable source such as the World Health Organization, a government health department, or a global health tracking website...''. \end{quote} 

Furthermore, the responses outputted by GPT-4V tend to be excessively verbose. For instance, in the above response, GPT-4V initially strays away from the user question by providing an unnecessary and potentially redundant description of the choropleth map. On average, GPT-4V responses were longer by 97.69 words, which can detract from user readability and concision. Yet, GPT-4V often performed reasonably well on certain visual questions such as asking overall shapes of trends (e.g., ``How is the graph moving?'' ``Would it be [...] going up like hills and valleys type of trend?'').


\section{Evaluation: User Study with Blind People}
\label{sec:user-study}
\begin{figure}
  \includegraphics[width=0.95\linewidth]{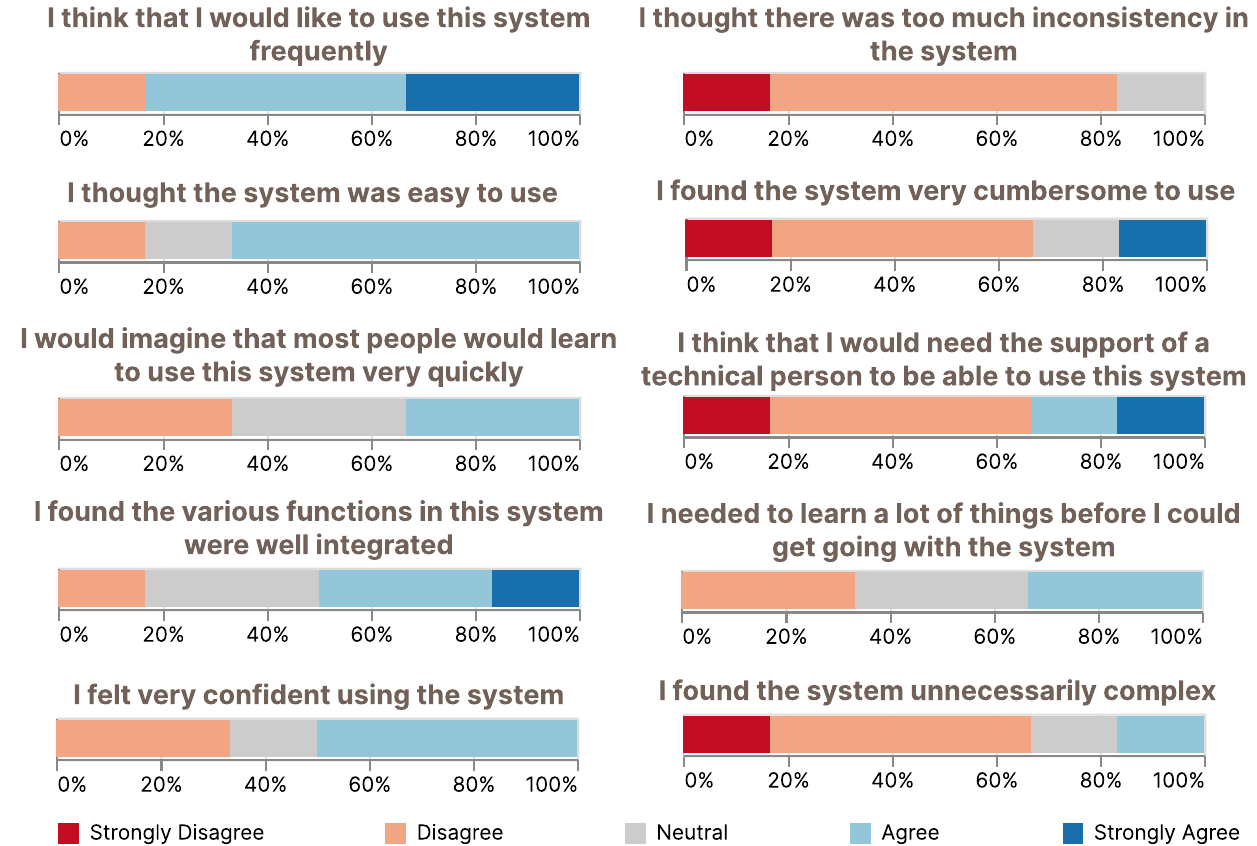}
  \caption{System Usability Scale Survey Results, showing participant responses on a Likert scale from `Strongly Disagree' to `Strongly Agree' for various statements about the system's frequency of use, ease of use, learning curve, integration of functions, consistency, need for technical support, and complexity, as part of the qualitative study.}
  \Description{The image displays a set of horizontal bar graphs depicting survey results on system usability, with responses ranging from 'Strongly Disagree' to 'Strongly Agree' for statements about frequency of use, ease of use, learning curve, integration of functions, user confidence, system consistency, complexity, and the need for technical support.}
  \label{fig:survey-result}
\end{figure}

 During the development process, we engaged with a blind participant who had prior experience using a screen reader on a daily basis. This participant provided feedback at two intermediate stages of development in what was an iterative design process.
 In addition to this intermediate prototype evaluation, we conducted a preliminary usability study with six additional blind/low-vision individuals, during which we prioritized self-guided investigation over pre-formed questions to simulate real-world encounters with charts. IRB approval was obtained before the commencement of the study. The stimuli used in the study was an early version of \toolname{}. While the major features remain unchanged, the improvements made after the study are described in Section \ref{sec:changes}.

\subsection{Participants}

\begin{table*}[]
\begin{tabular}{@{}lllllll@{}}
\toprule
PID & Gender & Age         & Vision Level           & Screen Reader Expertise                                                                  & Screen Reader & Chart Selected \\ \midrule
P1  & Male   & 45-54       & Blind with later onset & Expert                    & JAWS               & Bar Chart      \\
P2  & Female & 65 or older & Blind since birth      & Advanced & VoiceOver          & Line Chart     \\
P3  & Female & 25-34       & Blind with later onset & Intermediate   & JAWS               & Choropleth Map \\
P4  & Female & 25-34       & Blind since birth      & Advanced & JAWS               & Scatterplot    \\
P5  & Male   & 45-54       & Blind with later onset & Expert                    & JAWS               & Bar Chart      \\
P6  & Male   & 55-64       & Blind with later onset & Advanced & NVDA               & Choropleth Map \\ \bottomrule
\end{tabular}

\caption{\label{tbl-participants} Distribution of Participant Information, detailing gender, age range, level of vision impairment, screen reader expertise, preferred screen reader technology, and the type of chart selected for the study.}
\end{table*}

We recruited six blind/low-vision individuals from the National Institute of the Blind~\cite{NFB}. Their demographics are shown in \autoref{tbl-participants}. We tried to recruit diverse participants based on their gender and screen reader expertise. Our participants comprise three females and three males. Their ages were distributed as follows: two participants aged 25-34, two aged 45-54, one aged 55-64, and one aged 65 or older. Two participants have been blind since birth, while the other four experienced blindness onset later in life. Regarding their proficiency, all participants possessed at least intermediate experience, with three classified as advanced and two as experts. In terms of assistive technology, four participants primarily used JAWS as their screen reader, while the remaining two utilized VoiceOver and NVDA, respectively.

\subsection{Procedure}
The hour-long experiment was conducted over Zoom, and moderated by the first author. Upon entering the session, participants opened up our system in a web browser and chose a chart of their choice among the four options: line chart, bar chart, scatterplot, or choropleth map (see \autoref{tbl-participants} for participants' choice of charts). To mimic a real-world encounter with \toolname{}, we refrained from giving any contextual information about each chart, in order not to introduce any bias into the participant's selection. 

The study was divided into three parts. The first two - spanning roughly 20 minutes each - focused on the individual components of our multimodal approach---the keyboard-navigable tree view and the conversational module. The data table was not included in the study. In the assessment of each component, participants had 5-10 minutes to explore it freely, noting strengths and weaknesses. Subsequently, the component's function was explained, followed by a 10-minute guided exploration led by the moderator.

To conclude, each participant was asked a series of component-specific questions. For the keyboard-navigable tree view, these included: ``Describe the chart to the best of your capabilities'', ``How easy was it to navigate this interface?'', ``How useful is this tool?'' For the conversational module, questions were: ``Describe the chart to the best of your capabilities'', ``Assign a ranking for each question type based on their usefulness.'', ``How easy was it to use this interface?'', ``How useful is this tool?'' 

The final part centered on the components' combined functionality to assess the potential advantages of their collaborative operation. After having been exposed to the entire system, participants were asked to reassess \toolname{} based on ease of use and functionality. We maintained a consistent order for the parts without randomization. We conducted a brief post-task survey, which we distributed to each participant immediately after their completion of the study, to learn about the participants' overall experience with \toolname{}. Participants completed their surveys within 24 hours of participating in the study, allowing sufficient time for reflection whilst ensuring that \toolname{} remained fresh in their minds.

\subsection{Analysis Process}
Our analysis process involved a qualitative method, considering the study's small scale. 
During each session, the session moderator took detailed notes, capturing key observations and participant responses in real-time. Post-session, they reviewed the recordings of the interviews and interactions. This step involved careful, repeated listening to the audio to extract in-depth insights and to cross-reference them with the notes taken during the sessions as well as usability survey results. The analysis focused on identifying recurring themes, patterns of user behavior, and specific instances of user-system interaction challenges or successes. The analytical process was iterative, with findings from the initial sessions informing subsequent reviews. Analysis results were discussed with a senior author to validate interpretations and ensure a diverse perspective in the analysis.

\subsection{Behavioral Observations}
Here, we detail participants' actions and feedback while using \toolname{} during the study sessions.

\subsubsection{Navigating the tree view}
Participants were able to utilize the tree view using arrow keys and tab shortcuts as reported in prior studies~\cite{zong2022rich,kim2023beyond}, although the learning curve proved to be slightly steeper for P2 and P5. P5 remarked on the ``cumbersome'' structure of the tree for the bar chart, noting that it was due to the presence of over 170 unique data values. Rather than tediously navigating through the data using the down arrow key, P5 wished for a more efficient method to move between specific nodes within the tree view. P2 echoed this sentiment, highlighting the risk of disorientation, particularly with larger and more intricate data sets.

Several participants (P1, P3, P4, P5, P6) independently recognized the distinctive structure of the tree view, which presents a data set through visual encoding variables. For example, P5, after navigating a choropleth map and expressing frustration over manually sifting through 172 countries without an apparent order, was pleasantly surprised when using the right arrow key led him to the same data set, this time organized by vaccination rates in 10 percent increments. This participant then confirmed that the tree view was more effective in conveying a visualization's structure compared to a traditional data table.

After having used their keyboard to navigate through the tree view, participants were asked to describe the visual stimuli to the best of their capabilities. Responses were mixed, with two participants (P3 and P4) only being able to identify the variables represented by each axis (country v. percent of population vaccinated, and average life expectancy v. GDP per capita, respectively). This result suggests that despite being a good overall indicator of chart structure, the tree view alone is not sufficient for complete data visualization. The result was reaffirmed by the usefulness rating most individuals attributed to the system, with the average hovering around a 3 out of 5. 


\subsubsection{Exploring the conversational module}

Although 4 Participants (P1, P2, P3, P5) gravitated towards the text input modality, all affirmed the importance of retaining an option for voice input as well. All but one participant (P1, P2, P3, P4, P5) immediately asked data-driven questions (either simple fetches for data, like ``What is the vaccination percentage for Haiti'' or more complex queries involving multiple steps), with P6 instead asking a contextual question: ''Is there a way to rank the various countries into continents?'' (regarding the choropleth map). This coincided with subsequent participant ratings for the usefulness of the four query types, with all users asserting \textsc{analytical} queries as the most useful for chart comprehension. Most users (P1, P2, P3, P5) could not fathom the possibility that more broad questions were supported.

Following this independent exploration of the conversational model, participants were made aware of the four distinct types of queries and were once again directed to input their own questions; however, this time around, they had to broadly adhere to one of the four query classifications. Users demonstrated a greater proficiency with the conversational module during this guided exploration, with P1 even chaining multiple individual queries to arrive at a broader understanding of the chart. By consecutively asking ``What is the temperature for 2020?'' and ``What color is 2020?'', the participant was able to deduce that the color ‘red’ indicates positive temperature values. 

We also observed an affinity for contextual queries among the participant pool. One user (P4) who had little to no experience with map visualizations prior to the study asked: ``What is a choropleth map?'', to which the LLM outputted a correct response. However, when the same participant asked, ``What is a temporal polarity'' (pertaining to the bar chart), the LLM responded with a definition tied to linguistics. Although initially taken aback, the user acknowledged the possible ambiguities with the word ``temporal polarity'' (which has multiple meanings), and upon rephrasing her query to incorporate more precision, received a more accurate response. The participant attributed her realization to the \toolname{}’s justification (outputted alongside the response), which explicitly told her that it sourced its answer from the internet.

\subsubsection{Integrating the two components}
Participants were then introduced to navigation queries. We explained the purpose of these queries, emphasizing their role in wayfinding and orientation, and then allowed them to formulate their own navigation queries. All users concurred that these queries were essential for understanding the tree view (P6: "Maybe not as much as for smaller datasets, but I definitely see their use for complex data like this"), a sentiment echoed in the usefulness ratings they assigned to the integrated system. While previous Likert scale ratings for the individual components averaged around 3, after this introduction, participants consistently rated the complete system between 4 and 5, with 5 being extremely useful.


Most participants tended to input short and concise navigation queries. Rather than inputting ``How do I get from my current location to the percentage vaccinated value for Guam'', one user (P5) opted for the much simpler ``Take me to Guam''. Showcasing its conversational strengths, our model was able to precisely identify the starting as well as ending nodes from this colloquial text, yielding the instructions: ``Press the right arrow key. Press the down arrow key two times.''

\subsection{User Feedback and Reflection}

Participants completed a post-study questionnaire based on the System Usability Scale (see \autoref{fig:survey-result}). Notably, most participants (4 Agree; 1 Strongly Agree; 1 Disagree) concurred with the statement: ``I found the various functions in this system were well integrated.'' Results can be found in \autoref{fig:survey-result}. Participants also valued \toolname{}'s commitment to accessibility and transparency, especially within the conversational module. They envisioned real-world applications for \toolname{}, relating it to their personal experiences. For instance, P1 saw its potential in providing testing accommodations for GRE exams, noting its superiority over human proctors in translating visual graphs. P6, who teaches the NVDA screen reader to the BLV community, expressed interest in incorporating the system into his lessons. However, there was also constructive feedback.

Although most participants deemed the structure of navigation query responses (a sequence of directions) to be satisfactory, P2 advised that the system should automatically transport the user’s cursor to the desired location, as opposed to currently requiring the user to manually traverse the tree view themselves. One participant (P5) sought more control over the nature of LLM responses outputted by the conversational model. He brought up the necessity of having some implementation of a dial to regulate the verboseness of the outputted answers. The same user who commented on the cumbersome structure of the tree view (P5) further elaborated that he would prefer a more concise raw data table in its place, especially for less extensive datasets. Apropos implementing a raw data table, P5 remarked: "I would prefer a simple table. I'm used to it. I know how to do it. Not all blind people do, but I do".

\subsection{Changes After User Feedback}
\label{sec:changes}

Here, we briefly describe how we incorporated the lessons learned from the user study.

\paragraph{Injecting chart information into contextual queries}

\hlmaroon{D3} Participant 4's experience revealed a limitation in our initial contextual query handling: \toolname{} often misinterpreted user intent due to missing chart information. For instance, it treated ``temporal polarity'' as a linguistic term, not a data dimension depicted in the chart. We addressed this by enriching prompts with chart tree view text (\autoref{fig:pipeline}), which led to correct interpretation of the data (e.g., "temporal polarity" in global temperature data). This observation also led to exploring query ambiguity in general, resulting in pre-processing question refinement (Section \ref{sec:mitigation}).

\paragraph{Automating navigation to end-points in the tree view} 

\hlmaroon{D4} Participant 2's preference for automatic traversal suggested that manually pressing a series of keys can be cumbersome, especially for lengthy navigation paths. Originally, we aimed to ensure transparency and grant users control over the process. However, recognizing this issue, we decided to introduce an option for automatic traversal (see Section \ref{sec:navigation-queries}). While manual navigation might become tedious with familiarity, we kept it as a default option to accommodate varying technical proficiencies.

\paragraph{Providing a raw data table} 

\hlgold{D5} To address Participant 5's concerns regarding the complexity of navigating the tree view, we introduced a conventional raw data table as an alternative (see \ref{sec:data-table}). Though this addition may not significantly contribute to the novelty of our work, it underscores our dedication to creating an inclusive system. 



\paragraph{Providing query suggestions} 

\hlmaroon{D6} The initial self-guided exploration of charts showed that most participants (P1, P2, P3, P5) struggled to ask questions beyond data-retrieval queries. Despite recognizing the value of visual, contextual, and navigation queries, participants were unaware of these query types until they were explicitly explained by the moderator. This observation, along with P2's suggestion for help documentation and preference for interactive guidance, led to the addition of query suggestions to the system (see Section \ref{sec:mitigation} and Section \ref{sec:user-experience}).




\section{Discussion \& Future Work}
Our evaluation studies underscore the potential of \toolname{} and also pinpoint areas for enhancement. We reflect on the limitations and challenges, paving the way for future opportunities.

\subsection{Limitations and Opportunities}

\paragraph{Customizing verbosity levels} Despite our initial aim to offer concise and informative answers, P5's recommendation for user-adjustable response verbosity underscored the importance of user agency over designer-imposed settings. Given that speech is processed serially, the text length read by screen readers becomes a pivotal design consideration. This concern has been reiterated in prior research~\cite{zong2022rich,ault2002evaluation,w3ccompleximages,jung2021communicating, customization_is_key}. Similarly, offering users the capability to customize node descriptions in the tree view could prove advantageous.

\paragraph{Enhancing understanding of user context and question answerability} Our quantitative study results show that there is still an opportunity to improve the conversation module. These enhancements encompass recognizing unanswerable questions, effectively managing broad queries, and further refining the accuracy of analytical and visual query responses. Although the conversational module is not perfect in interpreting the ambiguous nature of natural languages, our efforts to make responses safe and explanatory still allowed participants to easily recover from mistakes.





\subsection{Need for Rigorous and Inclusive Benchmark Testing}
\label{sec:benchmark}
The cornerstone of our work is the conversational module, designed to address the challenges with keyboard navigation. While the existing dataset enabled a meaningful evaluation of response quality based on real-world queries, our study revealed the need for a more extensive and inclusive benchmarking dataset that incorporates viewpoints of blind and low-vision individuals~\cite{gadiraju2023wouldn}. 

\paragraph{Addressing advanced query types for a variety of chart types} Our evaluation was constrained not only by the four chart types but also by the limited range of questions (e.g., three query types), preventing a full assessment of \toolname{}'s scalability to larger datasets and more nuanced queries. For example, although we noted some visual queries necessitating reasoning over data and contextual queries seeking chart information, our limited dataset was not sufficient to capture multi-category questions requiring advanced multi-hop reasoning~\cite{chen2020hybridqa}. Furthermore, our study did not evaluate situational questions related to a user's current interest point within the tree view, which our dataset lacks. Questions dependent on understanding previous conversational context were also not explored. Considering the generative abilities of LLMs, synthetically generating these types of questions using human-created examples and advanced prompt engineering might be a viable method~\cite{ko2023natural}.

\paragraph{Incorporating vision and testing with varied metrics} 
Although GPT4V falls short in performance compared to \toolname{}, it shows promising visual description capabilities. This advancement contrasts with a few years ago when interpreting synthetic images like graphic designs and data visualizations was inferior to natural scene images~\cite{bylinskii2017learning}. Similarly, recent image-based ChartQAs still depend on OCR to extract text from images and create data tables~\cite{hoque2022chart,kantharaj2022opencqa}. Analyzing the capabilities of emerging vision-LLMs---from low-level analytic tasks like value look-ups and comparisons to higher-level cognitive operations like explaining chart patterns---will help find ways to integrate vision capability with the symbolic processing of \toolname{} to achieve greater performance. Moreover, expanding beyond mere correctness to include other pertinent measures such as fluency, informativeness, and relevance to the query~\cite{li2024leveraging} will be helpful for further improving the user experience of \toolname{}.







\subsection{Integrating into Existing Visualization Tools}

\paragraph{Accommodating practitioners' visualization workflows} Since \toolname{} operates under the assumption that a chart specification is available, it may not be directly applicable to charts currently found on the web. Instead, our vision is to integrate \toolname{} within existing data visualization platforms. Prior research underscores that many data visualization practitioners base their choices on the accessibility features of these platforms~\cite{joyner2022visualization}. Another study highlights the lack of accessible design support these tools offer~\cite{kim2023beyond}. Exploring the design space to determine how \toolname{} can seamlessly fit into current data visualization workflows would be compelling. Additionally, considering the degree of customization for data visualization designers, such as setting default verbosity levels and offering query guidance \& suggestions, warrants further investigation. 

\paragraph{Supporting multiple-coordinated and dynamic data visualizations} Furthermore, many existing visualization tools are evolving beyond creating singular charts to producing multi-coordinated charts or dashboards. Exploring the expansion of \toolname{} to accommodate these advanced forms of visualization presents an intriguing challenge. A recent study by Srinivasan et al.~\cite{srinivasan2023azimuth} investigates this problem by developing dashboards tailored for screen reader navigation, integrated with descriptions that aid in dashboard comprehension and interaction. Key questions arise in this context: How might we integrate a conversational agent into these dashboards? How should the agent resolve ambiguities in user queries about relevant charts? What constitutes an ideal design for mixed-initiative interaction in such environments? Furthermore, an exciting frontier is enabling users to interactively generate charts for data exploration. These questions open up new and unexplored avenues in the field of visualization accessibility.

\section{Conclusion}

In this study, we introduced \toolname{}, a tool that enhances structured chart navigation through conversational interactions. Our quantitative assessments demonstrate a notable advancement beyond existing systems, and our qualitative analyses underscore the importance of the system's integrated approach and its dedication to transparency. As a direction for future research, we aim to develop a more comprehensive and inclusive benchmark dataset to foster continuous enhancements of \toolname{}, along with integrating vision capabilities to bridge the current performance gaps.

\begin{acks}
The authors would like to thank Jacob Ottiger for initial guidance with \toolname{}'s system architecture, Sarah Zhang and Stephen Gwon for their work on synthesizing ground truths to dataset of BLV user queries, Nhung Van for helping develop the web application, and the reviewers for their insightful feedback. This research was supported in part by the Research Incentive Grant provided by Boston College.

\end{acks}

\bibliographystyle{ACM-Reference-Format}
\bibliography{bibliography}


\end{document}